\documentclass[lettersize,journal]{IEEEtran}
\usepackage{amsmath,amsfonts}
\usepackage{algorithmic}
\usepackage{algorithm}
\usepackage{array}
\usepackage{subeqnarray}
\usepackage{cases}
\usepackage{booktabs}
\usepackage{textcomp}
\usepackage{stfloats}
\usepackage{url}
\usepackage{verbatim}
\usepackage{graphicx}
\usepackage{cite}
\usepackage{float}
\usepackage {subfig}
\usepackage{gensymb}

\usepackage{hyperref}
\hypersetup{hypertex=true,
            colorlinks=true,
            linkcolor=blue,
            anchorcolor=blue,
            citecolor=blue}
            
\begin{document}

\title{Mainlobe Jamming Suppression Using MIMO-STCA Radar}




\author{\IEEEauthorblockN{
Huake Wang
\IEEEauthorrefmark{1}, ~\IEEEmembership{Member,~IEEE},
Bairui Cai,
Guisheng Liao, ~\IEEEmembership{Senior Member,~IEEE}}

\thanks{This work was supported in part by the National Natural Science Foundation of China under Grants 62301410, in part by the Aviation Science Foundation under Grants 20230001081021, in part by the National Key Laboratory of Air-based Information Perception and Fusion under Grants ZH 2024-0053, and in part by the National Natural Science Foundation of China under Grants 62331019.}
\thanks{Huake Wang, Yinghui Quan and Guisheng Liao are with School of Electronic Engineering,
Xidian University, Xi'an, Shaanxi, China, 710071. Huake Wang, Bairui Cai, Yinghui Quan and Guisheng Liao are with Hangzhou Institute of Technology, Xidian University, Hangzhou, Zhejiang, China, 311231.}
}




\maketitle

\begin{abstract}
Radar jamming suppression, particularly against mainlobe jamming, has become a critical focus in modern radar systems. This article investigates advanced mainlobe jamming suppression techniques utilizing a novel multiple-input multiple-output space-time coding array (MIMO-STCA) radar. Extending the capabilities of traditional MIMO radar, the MIMO-STCA framework introduces additional degrees of freedom (DoFs) in the range domain through the utilization of transmit time delays, offering enhanced resilience against interference. One of the key challenges in mainlobe jamming scenarios is the difficulty in obtaining interference-plus-noise samples that are free from target signal contamination. To address this, the study introduces a cumulative sampling-based non-homogeneous sample selection (CS-NHSS) algorithm to remove target-contaminated samples, ensuring accurate interference-plus-noise covariance matrix estimation and effective noise subspace separation. Building on this, the subsequent step is to apply the proposed noise subspace-based jamming mitigation (NSJM) algorithm, which leverages the orthogonality between noise and jamming subspace for effective jamming mitigation. 
However, NSJM performance can degrade due to spatial frequency mismatches caused by DoA or range quantization errors. To overcome this limitation, the study further proposes the robust jamming mitigation via noise subspace (RJNS) algorithm, incorporating adaptive beampattern control to achieve a flat-top mainlobe and broadened nulls, enhancing both anti-jamming effectiveness and robustness under non-ideal conditions. Simulation results verify the effectiveness of the proposed algorithms. Significant improvements in mainlobe jamming suppression are demonstrated through transmit-receive beampattern analysis and enhanced signal-to-interference-plus-noise ratio (SINR) curve.
\end{abstract}

\begin{IEEEkeywords}
Multiple-input multiple-output space-time coding array(MIMO-STCA), Non-homogeneous sample selection, Noise subspace, Beampattern synthesis, Robust jamming suppression.
\end{IEEEkeywords}

\section{Introduction}
\IEEEPARstart{W}{ith}  the rapid advancement of digital radio frequency memory[1] (DRFM) technology, the ability of jammers to generate active interference has severely compromised the target detection and parameter estimation capabilities of radar systems. Jamming can generally be classified into sidelobe and mainlobe jamming. Sidelobe jamming mitigation techniques, such as sidelobe cancellation (SLC) methods [2-4] and low sidelobe antenna designs [5-7], are well-established. However, mainlobe jamming poses a more significant challenge due to its inherent similarity to the true target’s antenna gain. Additionally, jammers replicate radar transmit signal at energy levels that exceed those of the true target, resulting in radar echoes dominated by jamming rather than target signals. This unique characteristic of mainlobe jamming, combined with deceptive jamming techniques, leads to a significant reduction in signal-to-interference-plus-noise ratio (SINR). The unique positioning of mainlobe jamming, coupled with deceptive jamming techniques, results in a notable decrease in SINR.

\begin{figure}[!t]
\centering
\includegraphics[width=2.5in]{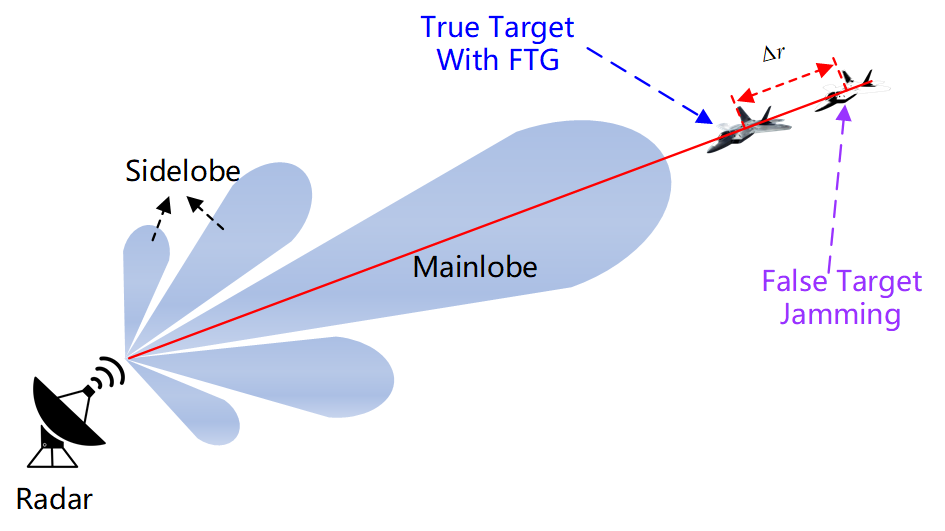}
\caption{Pictorial sketch of Mainlobe range deceptive jamming scenarios}
\label{fig_1}
\end{figure}

To suppress mainlobe jamming, researchers have explored various approaches across multiple domains, including the spatial domain [8-10], time domain [11-14], and frequency domain [15-17]. 

Scholars have extensively explored the use of angular differences between target and jamming signals for spatial domain jamming suppression. A classic approach involves adaptive beamforming to mitigate interference. In [8], blocking matrices are employed during the preprocessing stage to effectively suppress mainlobe jamming. Meanwhile, [9] enhances anti-jamming performance by optimizing mutual coupling compensation techniques. However, these methods often introduce unwanted beampattern distortion. To address this, the monopulse technique can be integrated with jamming suppression algorithms, as demonstrated in [10]. This combined approach utilizes array gain constraints and monopulse ratio curves to suppress multiple jamming sources while simultaneously minimizing beampattern distortion.

In the context of time-domain jamming suppression, the blind source separation (BSS) algorithm is particularly well-known. Recent work in [11] propose a multiple mainlobe jamming suppression technique using an eigen-projection-based BSS algorithm. In [12], a method involving four space-time sum-difference channels is formulated, leveraging two-dimensional space-time information to enhance the discriminability of source signals. This approach is combined with the joint approximate diagonalization of eigenmatrices (JADE) BSS algorithm to achieve effective jamming suppression. In [13], a novel method is introduced, presenting a parameter estimation approach for intra-pulse interrupted sampling repeater jamming (ISRJ). This technique estimates the sampling duration and repeat sampling interval using the Hilbert transform and peak detection techniques. Subsequently, the sampling number and forwarding times are determined, facilitating the design of jamming strategies adapted to different operational scenarios. Furthermore, [14] proposes using block cipher output as a phase code for pulse compression, effectively suppressing mainlobe jamming.

In the context of frequency-domain mainlobe jamming suppression, [15] and [16] propose an adaptive frequency agility mechanism based on the design principles of carrier frequency, significantly improving anti-jamming capabilities. The approach introduced in [17] employs an optimization-based method to design radar waveforms with phase agility between pulses, thereby enhancing the signal-to-noise ratio (SNR) for target detection. This is achieved by constructing Doppler frequency domain notches to counteract velocity deception jamming.

The utilization of orthogonal coding techniques between channels allows for the extraction of range dimension freedom at the receiving end. This feature enables STCA to effectively address the challenge of suppressing mainlobe range false target jamming[18]. The design of STCA transmit delay and receive filters, aimed at enhancing the performance of target detection in the presence of jamming, is investigated in [19]. Additionally, STCA achieves lower sidelobes by introducing phase coding between array elements[20]. [21] discusses the advantages of space-time coding technology in enhancing range resolution, reducing sidelobe levels, and improving beampattern design capabilities. Although research on mainlobe jamming suppression within the STCA systems remains limited, its excellent characteristics have significant potential for application in jamming mitigation.

Therefore, we focus on investigating adaptive jamming suppression algorithms based on STCA. The self-nulling phenomenon in signals can significantly degrade the performance of adaptive beamformers, as reported in [22]-[23]. The phenomenon of self-nulling in signals can substantially degrade the output performance of adaptive beamformers. Therefore, to improve the jamming suppression performance of adaptive beamforming, it is essential to acquire a relatively pure interference-plus-noise covariance matrix (INCM). Traditional methods include the inner product method [24] and the generalized inner product (GIP) algorithm [25]-[26]. These algorithms employ non-uniform detectors to select non-homogeneous samples and then constructs a pure INCM. In recent years, some scholars propose the reconstruction method of INCM to obtain a purer covariance matrix[27][28], thereby achieving a better SINR. In [29], the approach involves dynamically adjusting the uncertainty region of the interference direction and simplifying the power spectral density function, thus effectively reconstructing the INCM. Furthermore, [30] develops a simplified jamming power estimation algorithm and proposes a robust adaptive beamforming algorithm based on the INCM reconstruction. However, the design complexity would lead to a heavy computational burden [31]. To tackle this problem, we introduce a cumulative sampling-based non-homogeneous sample selection (CS-NHSS) algorithm to exclude target-contaminated samples.

In real-world scenarios, the parameters of actual targets and jamming signals may not align perfectly with theoretical models. Consequently, it is imperative to devise a robust jamming suppression algorithm for the STCA radar, enabling the radar system to effectively counteract jamming signals even under non-ideal conditions. Traditional robust jamming suppression algorithms include diagonal loading [32] and covariance matrix reconstruction methods [33]. [33] introduces a method that employs Capon power spectrum to construct a circular uncertainty set in the non-target angle region, and then reconstructs the INCM through integration to enhance jamming suppression capability under the condition of array steering vector mismatch. However, this method is computationally intensive due to the high redundancy. The methods based on convex optimization theory, such as successive quadratically constrained quadratic programming (QCQP) [34] and semi-definite relaxation (SDR) [35] techniques, are characterized by their high computational complexity and extended processing times. Additionally, the application of optimization algorithms like genetic algorithms [36] and ant colony optimization [37] to robust pattern design has been explored, yet these heuristic approaches lack the precision control for certain applications. 

 In recent years, there has been a growing interest in methods based on adaptive array response control due to their precise control and computational flexibility. [38] investigates single-point pattern control, while [39] utilizes orthogonal projection operators to achieve pattern control. Building upon this, [40] enhance the algorithm’s efficiency by applying the maximum magnitude response principle. [41] expands pattern control with multiple virtual jammers, though it lacks a closed-form solution. [42] advances this by calculating optimal weights for enhanced beampattern flexibility. It is noteworthy that this paper introduces adaptive array beampattern control to enhance the robustness of the interference suppression performance for MIMO-STCA radar systems. 
 
In the pursuit of radar jamming suppression, with a focus on countering mainlobe jamming, this article presents a pivotal advancement in modern radar system technology. The contributions of this study are outlined as follows:

1) We investigate advanced mainlobe jamming suppression techniques using a novel Multiple-Input Multiple-Output Space-Time Coding Array (MIMO-STCA) radar. This framework extends traditional MIMO radar by introducing time delays between transmit elements. Extra degrees of freedom (DoFs) in the range domain can be extracted to mitigate the mainlobe range false target jamming.

2) We introduce a cumulative sampling-based non-homogeneous sample selection (CS-NHSS) algorithm to obtain interference-plus-noise samples free from target signal contamination. This algorithm effectively excludes target-contaminated samples, ensuring precise estimation of the interference-plus-noise covariance matrix and facilitating effective noise subspace separation.

3) To address the challenge of mainlobe interference suppression, we propose a noise subspace-based interference mitigation (NSJM) algorithm that exploits the orthogonality between the noise and jamming subspaces to achieve effective jamming suppression. 

4) To mitigate the impact of system errors on performance, we develop the robust jamming mitigation via noise subspace (RJNS) technique. This approach can realize adaptive beampattern control to create a flat mainlobe and wide nulls, improving anti-jamming robustness in practical applications.

The rest of this article is organized as follows. The signal model of MIMO-STCA is introduced in Section II. In Section III, the CS-NHSS and NSJM algorithms are introduced in detail. Moreover, the RJNS al are presented in Section IV, and the algorithm analysis is described. Section V  conducts simulation verification of the proposed methods. Finally, the conclusion is given in Section VI.

\section{SIGNAL MODEL of MIMO-STCA RADAR}
\subsection{Target Signal Model}

Let us consider a colocated MIMO configuration consisting of transmit elements and receive elements with uniform spacing $d$ ($d$ is set to half wavelength). A small time delay $\Delta t$ is introduced between the transmit elements, with the first element selected as the reference element. The signal model is shown in Fig.2 .

\begin{figure}[!h]
\centering
\includegraphics[width=2.5in]{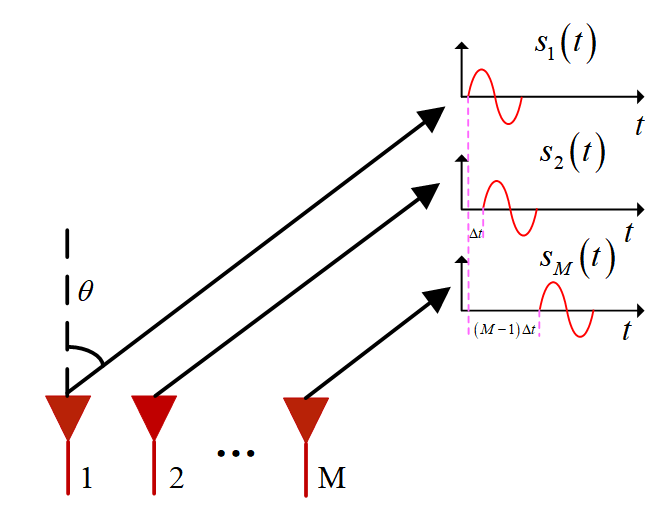}
\caption{MIMO-STCA signal model}
\label{fig_1}
\end{figure}

The small time delay introduced by the $m$-th transmit element can be expressed as:

\begin{equation}
t_{m}=(m-1) \Delta t, m=1,2, \cdots, M
\end{equation}

Thus, the transmission signal $s_{m}(t)$  of the $m$-th transmitting array element can be represented as:

\begin{equation}
\begin{array}{c}s_{m}(t)=e^{j 2 \pi f_{0} t} \varphi_{m}\left(t-t_{m}\right)
\end{array}
\end{equation}

MIMO-STCA radar combines the characteristics of MIMO radar, where the waveforms transmitted between different array elements are mutually orthogonal, satisfying the orthogonality definition as follows:

\begin{equation}
\int_{T_{p}} \varphi_{m}(t) \varphi_{m^{\prime}}(t) d t=0, \forall m, m^{\prime}=1,2, \ldots, M, m \neq m^{\prime}
\end{equation}
where $T_{p}$ is the radar pulse duration, $\varphi_{m}(t)=c_{m} g(t)$ represents the orthogonal signal transmitted by the $m$-th array element, $c_{m}$ stands for the orthogonal encoding signal, and

\begin{equation}
\begin{array}{c}g(t)=\operatorname{rect}\left(\frac{t}{T_{p}}\right) \exp \left(\mathrm{j} \pi \mu t^{2}\right)
\end{array}
\end{equation}
where $rect\left( t \right) = \left\{ {\begin{array}{*{20}{c}}
{1,}&{0 \le t \le 1}\\
{0,}&{else}
\end{array}} \right.$
denotes the chirp rate.

Assuming a target is located at position $(\theta_{0},R_{0})$, the reflected signal received by the $n$-th receiving array element can be represented as: 

\begin{equation}
\begin{array}{l}
{x_n}\left( {t,{\theta _0}} \right)\\
{\rm{ = }}{e^{j2\pi \frac{d}{\lambda }\left( {n - 1} \right)\sin \left( {{\theta _0}} \right)}}\sum\limits_{m = 1}^M {[{\varphi _m}\left( {t - 2{\tau _0} - {\tau _m} - {\tau _n}} \right)} \\\hspace{0.2cm}
{\rm{                            }} \cdot {e^{j2\pi \frac{d}{\lambda }\left( {m - 1} \right)\sin \left( {{\theta _0}} \right)}}{e^{ - j\pi \mu 2\left( {t - 2{\tau _0}} \right)\left( {m - 1} \right)\Delta t}}{e^{j2\pi {f_0}\left( {t - 2{\tau _0} - {\tau _m} - {\tau _n}} \right)}}]
\end{array}\
\end{equation}
where $\tau_{m}=\frac{(m-1) d \sin \left(\theta_{0}\right)}{c}, \tau_{n}=\frac{(n-1) d \sin \left(\theta_{0}\right)}{c}, \tau_{0}=\frac{R_{0}}{c}$ and $c$ being the wave path differences in transmit and receive arrays, the reference time delay and the speed of light, respectively.

The above expression can be rewritten in the narrowband condition as:

\begin{equation}
\begin{array}{l}
{x_n}\left( {t,{\theta _0}} \right)\\
{\rm{ = }}{e^{j2\pi {f_0}\left( {t - 2{\tau _0}} \right)}}{e^{j2\pi \frac{d}{\lambda }\left( {n - 1} \right)\sin \left( {{\theta _0}} \right)}}\sum\limits_{m = 1}^M {[{\varphi _m}\left( {t - 2{\tau _0}} \right)} \\
{\rm{                                          }} \cdot {e^{j2\pi \frac{d}{\lambda }\left( {m - 1} \right)\sin \left( {{\theta _0}} \right)}}{e^{ - j\pi \mu 2\left( {t - 2{\tau _0}} \right)\left( {m - 1} \right)\Delta t}}]
\end{array}\
\end{equation}

After down-conversion, the signal can be expressed as

\begin{equation}
\begin{array}{l}
{{\mathord{\buildrel{\lower3pt\hbox{$\scriptscriptstyle\smile$}} 
\over x} }_n}\left( {t,{\theta _0}} \right)\\
{\rm{ = }}\beta {e^{j2\pi \frac{d}{\lambda }\left( {n - 1} \right)\sin \left( {{\theta _0}} \right)}}\sum\limits_{m = 1}^M {[{\varphi _m}\left( {t - 2{\tau _0}} \right)} \\
{\rm{                               }} \cdot {e^{j2\pi \frac{d}{\lambda }\left( {m - 1} \right)\sin \left( {{\theta _0}} \right)}}{e^{ - j\pi \mu 2\left( {t - 2{\tau _0}} \right)\left( {m - 1} \right)\Delta t}}]
\end{array}\
\end{equation}
where $\beta=e^-j 2 \pi f_{0}2\tau_{0}$ is the complex echo amplitude. 

After the digitally mixed related to $\Delta t$, the processed echo in the $m$-th  channel can be expressed as

\begin{equation}
\begin{array}{l}
{{\mathord{\buildrel{\lower3pt\hbox{$\scriptscriptstyle\smile$}} 
\over x} }_{n,m}}\left( {t,{\theta _0}} \right)\\
{\rm{ =  }}{{\mathord{\buildrel{\lower3pt\hbox{$\scriptscriptstyle\smile$}} 
\over x} }_n}\left( {t,{\theta _0}} \right) \cdot {e^{j2\pi \mu t\left( {m - 1} \right)\Delta t}}\\
 = \beta {e^{j2\pi \frac{d}{\lambda }\left( {n - 1} \right)\sin \left( {{\theta _0}} \right)}}{\varphi _m}\left( {t - 2{\tau _0}} \right){e^{j2\pi \frac{d}{\lambda }\left( {m - 1} \right)\sin \left( {{\theta _0}} \right)}}{e^{j2\pi \mu 2{\tau _0}\left( {m - 1} \right)\Delta t}}\\\hspace{0.2cm}
{\rm{  }} + \beta {e^{j2\pi \frac{d}{\lambda }\left( {n - 1} \right)\sin \left( {{\theta _0}} \right)}}\sum\limits_{l = 1,l \ne m}^M {[{\varphi _l}\left( {t - 2{\tau _0}} \right){e^{j2\pi \frac{d}{\lambda }\left( {l - 1} \right)\sin \left( {{\theta _0}} \right)}}} \\\hspace{0.2cm}
{\rm{                                        }} \cdot {e^{j2\pi \mu 2{\tau _0}\left( {l - 1} \right)\Delta t}}{e^{j2\pi \mu t\left( {l - 1} \right)\Delta t}}]
\end{array}\
\end{equation}

Further, the orthogonal matched filter is designed for the $m$-th transmit waveform as follows

\begin{equation}
\begin{array}{c}h_{m}(t)=\varphi_{m}^{*}(-t)
\end{array}
\end{equation}

Hence, the result of $\breve{x}_{n, m}\left(t, \theta_{0}\right)$  after a bank of matched filter can be expressed as

\begin{equation}
\begin{array}{l}
{{\hat x}_{n,m}}\left( {t,{\theta _0}} \right)\\
 = {{\mathord{\buildrel{\lower3pt\hbox{$\scriptscriptstyle\smile$}} 
\over x} }_{n,m}}\left( {t,{\theta _0}} \right)*{h_m}\left( t \right)\\
 = \beta {e^{j2\pi \frac{d}{\lambda }(n - 1)\sin {\theta _0}}}{e^{j2\pi \frac{d}{\lambda }(m - 1)\sin {\theta _0}}}{e^{j2\pi \mu {\rm{2}}{\tau _{\rm{0}}}(m - 1)\Delta t}}\\
 \hspace{0.2cm}+ \beta {e^{j2\pi \frac{d}{\lambda }(n - 1)\sin {\theta _0}}}\sum\limits_{l = 1,l \ne m}^M {[{e^{j2\pi \frac{d}{\lambda }(l - 1)\sin {\theta _0}}}} \\
\hspace{0.2cm}\cdot {e^{j2\pi \mu {\rm{2}}{\tau _{\rm{0}}}(l - 1)\Delta t}}\int {{\varphi _l}} \left( {\tau  - {\tau _0}} \right)\varphi _m^*(\tau  - t){e^{j2\pi \mu t(m - l)\Delta t}}d\tau ]
\end{array}
\end{equation}
where ‘*’ indicates convolution operation, ‘$(\cdot)^{*}$’denotes conjugate operation."

Due to $\varphi_{m}(t)$ satisfy the orthogonality condition, the above expression can be further represented as:

\begin{equation}
\begin{array}{l}
{{\hat x}_{n,m}}\left( {t,{\theta _0}} \right)\\
 = \beta {e^{j2\pi \frac{d}{\lambda }(n - 1)\sin {\theta _0}}}{e^{j2\pi \frac{d}{\lambda }(m - 1)\sin {\theta _0}}}{e^{j2\pi \mu 2{\tau _0}(m - 1)\Delta t}}
\end{array}
\end{equation}

Consequently, the output $\mathbf{x}_{n}\left(t, \theta_{0}\right)$ of the $M$-dimensional matched filter for the target echo received by the -th receiving array element is:

\begin{equation}
\begin{array}{c}\mathbf{x}_{n}\left(t, \theta_{0}\right)=\left\lceil\hat{x}_{n, 1}\left(t, \theta_{0}\right), \hat{x}_{n, 2}\left(t, \theta_{0}\right), \cdots, \hat{x}_{n, M}\left(t, \theta_{0}\right)\right\rceil^{T}
\end{array}
\end{equation}

Finally, with $N$ receive channels, the total received signals of the target in MIMO-STCA can be represented as a column vector of dimension $MN\times1$. The specific form of the vector is given by:

\begin{equation}
\begin{array}{c}\begin{aligned} \mathbf{x}_{s} & =\left[\mathbf{x}_{1}\left(t, \theta_{0}\right), \mathbf{x}_{2}\left(t, \theta_{0}\right), \cdots, \mathbf{x}_{N}\left(t, \theta_{0}\right)\right]^{T} \\ & =\beta \mathbf{b}\left(\theta_{0}\right) \otimes \mathbf{a}\left(R_{0}, \theta_{0}\right)\end{aligned}
\end{array}
\label{eq13}
\end{equation}
where $\otimes$ represents the Kronecker product, ${\bf{b}}\left( \theta_0 \right)$ is the receive steering vector, 
${\bf{a}}\left( R_0, \theta_0 \right)$ is the transmit steering vector, and their expressions are as follows:

\begin{equation}
\begin{array}{c}
\begin{aligned} 
{\bf{b}}\left( {{\theta _0}} \right) = {\left[ {1,{e^{j2\pi \frac{d}{\lambda }\sin {\theta _0}}}, \cdots ,{e^{j2\pi \frac{d}{\lambda }\left( {N - 1} \right)\sin {\theta _0}}}} \right]^{\rm{T}}}
\end{aligned}
\end{array}
\end{equation}

\begin{equation}
\begin{array}{l}
\begin{aligned} 
\begin{split}
{\bf{a}}\left( R_0, \theta_0 \right) = \bigl[ & 1, e^{j2\pi \frac{d}{\lambda} \sin \theta_0} e^{j4\pi \mu \tau_0 \Delta t}, \cdots, \\
& e^{j2\pi \frac{d}{\lambda} (M - 1) \sin \theta_0} e^{j4\pi \mu \tau_0 (M - 1) \Delta t} \bigr]
\end{split}
\end{aligned}
\end{array}
\end{equation}

\subsection{False target signal model}

Assuming the jammer which is false target generator (FTG) and located at $(\theta_{j},R_{j})$, intercepts our radar signal and then performs a storage-and-forward operation. For mainlobe jamming, the angle $\theta_{j}$ of the jammer is equal to $\theta_{0}$.

The radar signal intercepted by the FTG is the sum of signals transmitted by each element of the radar array at location $(\theta_{j},R_{j})$, and can be represented as:

\begin{equation}
\begin{array}{l}
s\left( {t,{\theta _j}} \right)\\
{\rm{ = }}\sum\limits_{m = 1}^M {[{\varphi _m}\left( {t - {\tau _m} - {\tau _j}} \right){e^{j2\pi \frac{d}{\lambda }\left( {m - 1} \right)\sin \left( {{\theta _j}} \right)}}} \\
 \hspace{3cm}\cdot {e^{ - j\pi \mu 2\left( {t - {\tau _j}} \right)\left( {m - 1} \right)\Delta t}}]
\end{array}\
\end{equation}
where $ \tau _ { j } = \frac { R _ { j } } { c } $ is equal to $ \tau _ { 0 } $.

The FTG can intercept and retransmit radar signals, effectively deceiving radar systems. Through proper timing control prior to the arrival of the next radar pulse, it creates false targets with range offsets that can be either positive or negative.

Assuming that the FTG forwards a total $Q$ false targets, for the $q$-th false target with a delay of  $\tau_{q,FTG}$, the forwarded signal can be represented as:

\begin{equation}
\begin{array}{l}
{s_q}\left( {t,{\theta _j}} \right)\\
{\rm{ = }}\sum\limits_{m = 1}^M {[{\varphi _m}\left( {t - {\tau _m} - {\tau _j} - {\tau _{q,FTG}}} \right){e^{j2\pi \frac{d}{\lambda }\left( {m - 1} \right)\sin \left( {{\theta _j}} \right)}}} \\
{\rm{\hspace{3cm}}} \cdot {e^{ - j\pi \mu 2\left( {t - {\tau _j} - {\tau _{q,FTG}}} \right)\left( {m - 1} \right)\Delta t}}]
\end{array}
\end{equation}

The signal of the $n$-th array element received from the FTG generated false target is the signal of the $q$-th false target, given by:

\begin{equation}
\begin{array}{l}
{x_{n,q}}\left( {t,{\theta _j}} \right)\\
{\rm{ = }}\sum\limits_{m = 1}^M {{\varphi _m}\left( {t - 2{\tau _j} - {\tau _m} - {\tau _n} - {\tau _{q,FTG}}} \right)} \\
 \hspace{1cm}\cdot {e^{j2\pi \frac{d}{\lambda }\left( {m - 1} \right)\sin \left( {{\theta _j}} \right)}}{e^{ - j\pi \mu 2\left( {t - 2{\tau _j} - {\tau _m} - {\tau _n} - {\tau _{q,FTG}}} \right)\left( {m - 1} \right)\Delta t}}\\
 \hspace{1cm}\cdot {e^{j2\pi \frac{d}{\lambda }\left( {n - 1} \right)\sin \left( {{\theta _0}} \right)}}{e^{j2\pi {f_0}\left( {t - 2{\tau _j} - {\tau _m} - {\tau _n} - {\tau _{q,FTG}}} \right)}}{\rm{          }}
\end{array}
\end{equation}

Under the narrowband assumption, while making $\tau_{q}=\tau_{j}+\frac{\tau_{q, F T G}}{2}$, the above equation can be rewritten as:

\begin{equation}
\begin{array}{l}
{x_{n,q}}\left( {t,{\theta _j}} \right)\\
{\rm{ = }}\sum\limits_{m = 1}^M {\left[ {{\varphi _m}\left( {t - {\rm{2}}{\tau _q}} \right){e^{j2\pi \frac{d}{\lambda }\left( {m - 1} \right)\sin \left( {{\theta _j}} \right)}}} \right.} \\
\left. {{\rm{      \hspace{1cm}   }} \cdot {e^{ - j\pi \mu 2\left( {t - {\rm{2}}{\tau _q}} \right)\left( {m - 1} \right)\Delta t}}{e^{j2\pi \frac{d}{\lambda }\left( {n - 1} \right)\sin \left( {{\theta _j}} \right)}}{e^{j2\pi {f_0}\left( {t - {\rm{2}}{\tau _q}} \right)}}} \right]
\end{array}
\end{equation}

After digital down-conversion and orthogonal matched filtering, the  $q$-th jamming generated by FTG can be represented as

\begin{equation}
\begin{array}{l}
{{\hat x}_{n,m,q}}\left( {t,{\theta _j}} \right)\\
 = \beta {e^{j2\pi \frac{d}{\lambda }(n - 1)\sin {\theta _j}}}{e^{j2\pi \frac{d}{\lambda }(m - 1)\sin {\theta _j}}}{e^{j2\pi \mu {\rm{2}}{\tau _q}(m - 1)\Delta t}}
\end{array}
\end{equation}

Thus, the output $\mathbf{x}_{n, q}\left(t, \theta_{i}\right)$ of the M-dimensional matched filter for the echo received by the $n$-th receiving array element can be expressed as:

\begin{equation}
\begin{array}{c}
\mathbf{x}_{n, q}\left(t, \theta_{j}\right)=\left[\hat{x}_{n, 1, q}\left(t, \theta_{i}\right), \hat{x}_{n, 2, q}\left(t, \theta_{i}\right), \cdots, \hat{x}_{n, M, q}\left(t, \theta_{i}\right)\right]^{T}
\end{array}
\end{equation}

Finally, with $N$ receive channels, the received signals of the $q$-th jamming can be represented as a column vector of dimension $MN\times1$. Its specific form is:

\begin{equation}
\begin{array}{c}
\begin{aligned} \mathbf{x}_{q} & =\left[\mathbf{x}_{1, q}\left(t, \theta_{j}\right), \mathbf{x}_{2, q}\left(t, \theta_{j}\right), \cdots, \mathbf{x}_{N, q}\left(t, \theta_{j}\right)\right]^{T} \\ & =\beta \mathbf{b}\left(\theta_{j}\right) \otimes \mathbf{a}\left(R_{q}, \theta_{j}\right)\end{aligned}
\end{array}
\end{equation}
where $R_{q}$ represents the modulation range of the false target generated by the FTG, and its modulated equivalent range is:

\begin{equation}
\begin{array}{c}
R_{q}=c \cdot \tau_{q}=c \cdot\left(\tau_{j}+\frac{\tau_{q, F T G}}{2}\right)
\end{array}
\end{equation}

Therefore, the modulation range of false target jamming is simultaneously related to the location of the FTG and the time delay of the modulation.

In conclusion, the signal received by MIMO-STCA can be expressed as:

\begin{equation}
\begin{array}{c}
{\bf{x}} = \sigma _s^2{{\bf{x}}_s} + \sum\limits_{q = 1}^Q {\sigma _q^2{{\bf{x}}_q}} {\rm{ +  }}{\bf{n}}
\end{array}
\end{equation}
where $\bf{n}$ is noise, $  \sigma _ { s } ^ { 2 }$ and $ \sigma _ { q } ^ { 2 }$ represent the power of the target and jammings respectively.

\section{Noise Subspace Jamming Mitigation}

The key challenge in noise subspace jamming suppression is to obtain a pure interference-plus-noise covariance matrix and a clean noise subspace.  Leveraging prior information, suitable training samples from the received echoes in MIMO-STCA systems can be chosen to suppress the target signal effectively.

\subsection{Cumulative Sampling-based non-homogeneous sample selection (CS-NHSS) algorithm}

The signals received by the MIMO-STCA radar consist of target or jamming signals and noise. A covariance matrix is constructed from these signals and subjected to eigenvalue decomposition, with the eigenvectors arranged in descending order according to their eigenvalues. If there are a total of  $Q$ mainlobe jammings in the echo signals, the covariance matrix $\mathbf{R}_{x}$ can be decomposed as follows:

\begin{equation}
\begin{array}{l}
{{\bf{R}}_x} = E\left[ {{\bf{x}}{{\bf{x}}^{\rm{H}}}} \right] = \sum\limits_{k = 1}^{MN} {{\lambda _k}} {{\bf{u}}_k}{\bf{u}}_k^{\rm{H}}\\
  \hspace{0.56cm} = \sum\limits_{k = 1}^Q {{\lambda _{jq}}{{\bf{u}}_{jq}}{\bf{u}}_{jq}^{\rm{H}}}  + {\lambda _s}{{\bf{u}}_s}{\bf{u}}_s^{\rm{H}} + \sum\limits_{k = s + 1}^{MN} {{\lambda _k}} {{\bf{u}}_k}{\bf{u}}_k^{\rm{H}}\\
{\lambda _{j{\rm{1}}}} \ge {\lambda _{j2}} \ge  \cdots {\lambda _{jQ}} \ge {\lambda _s} \ge {\lambda _{s + 1}} \cdots  = {\lambda _{MN}}
\end{array}
\label{eq23}
\end{equation}
where $\mathbf{x}$ represents the received data echoes, denotes the conjugate transpose, $\lambda_{k}$ and $\mathbf{u}_{k}$ represent the eigenvalues and eigenvectors obtained after decomposing the covariance matrix, $\lambda_{j}$ and $\mathbf{u}_{j}$ represents the eigenvalue and eigenvectors of the jamming signal, $\lambda_{s}$ and $\mathbf{u}_{s}$ represents the eigenvalue and eigenvectors of the true target. 

By utilizing the expected target steering vector 
${{\bf{v}}_{s0}} = {\bf{b}}({\theta _{s0}}) \otimes {\bf{a}}({R_{s0}},{\theta _{s0}})$, which is consist of expected target receive steering vector and transmit steering vector. The location $ \rho$ and correlation coefficient $ \gamma$ of the eigenvalue corresponding to the eigenvector with the strongest correlation to $\mathbf{v}_{s0}$ can be determined.

Under ideal conditions, the strongest correlation exists between the expected target steering vector $\mathbf{v}_{s0}$ and the eigenvector corresponding to the true target. This feature can be utilized to ascertain the presence of a true target in the received signal and return the respective eigenvector of the target along with its position in the eigenvector sequence.

\begin{equation}
\begin{array}{c}\left[ \mathbf{u},\gamma,\rho\right]=\arg \max \left|\mathbf{u}_{k}^{\mathrm{H}} \mathbf{v}_{s 0}\right|^{2}
\end{array} 
\label{eq24}
\end{equation}

Specially, the $\rho$ not only can indicate the position of the eigenvalue with the strongest correlation to the desired target steering vector in the descending order, but also can reflect the existence of $(\rho-1)$ false target jammings in this signal.

Then, setting a sampling threshold $\eta$, and data points above ${{\bf{x}}_{jq}}$ the threshold are selected. Since some noise energy may exceed the threshold, consecutive data points surpassing threshold and covering more than 4/5 of the signal pulse width can be chosen as cumulative samples. 

The data satisfying this threshold condition are considered as the target signal or jamming signal. Let ${X}_{\mathbf{Q}}$ denote the sampled data meeting this criterion.

\begin{equation}
\begin{array}{c}
\mathbf{X}_{\mathbf{Q}}=\left[\mathbf{x}_{j 1}, \mathbf{x}_{j 2}, \cdots, \mathbf{x}_{j q}, \cdots, \mathbf{x}_{j Q}\right], \mathbf{x}_{j q} \geq \eta \quad q \in[1: Q]
\end{array}
\end{equation}

Construct the covariance matrix from the sampled data $\mathbf{X}_{\mathbf{Q}}$ , and calculate the maximum correlation coefficient with the desired target steering vector using Equation(\ref{eq24}). Set a threshold $\chi$ to determine the presence of the target. If no target is detected, the sampled data $\mathbf{X}_{\mathbf{Q}}$ can be utilized as training samples to form the adaptive weight vector.

\begin{equation}
\begin{array}{c}
\left\{\begin{array}{l}\mathrm{H}_{0}: \gamma<\chi \\ \mathrm{H}_{1}: \gamma \geq \chi\end{array}\right.
\end{array}
\end{equation}
where $\chi$ is judgment threshold. If $\gamma < \chi$, it indicates the absence of a target. Conversely, if the correlation coefficient $\gamma$ is greater than or equal to $\chi$ , the target is deemed to exist. In this case, the approximate position of the target in the time domain should be analyzed using cumulative sampling, and the received signals in the array should be selectively sampled cumulatively.

In the H1 condition, signals exceeding the threshold $\eta $ are sampled cumulatively, yielding sampled data $
{{\bf{X}}_q} = [{{\bf{x}}_{j1}},{{\bf{x}}_{j2}},...{{\bf{x}}_{jq}}]$ from the $q$-th cumulative sampling, arranged from near to far. These sampled signals are then used to construct the $q$-th covariance matrix ${{\bf{R}}_{{\bf{X}}q}}$ and perform eigenvalue decomposition.

\begin{figure}[!t]
\centering
\includegraphics[scale=0.12]{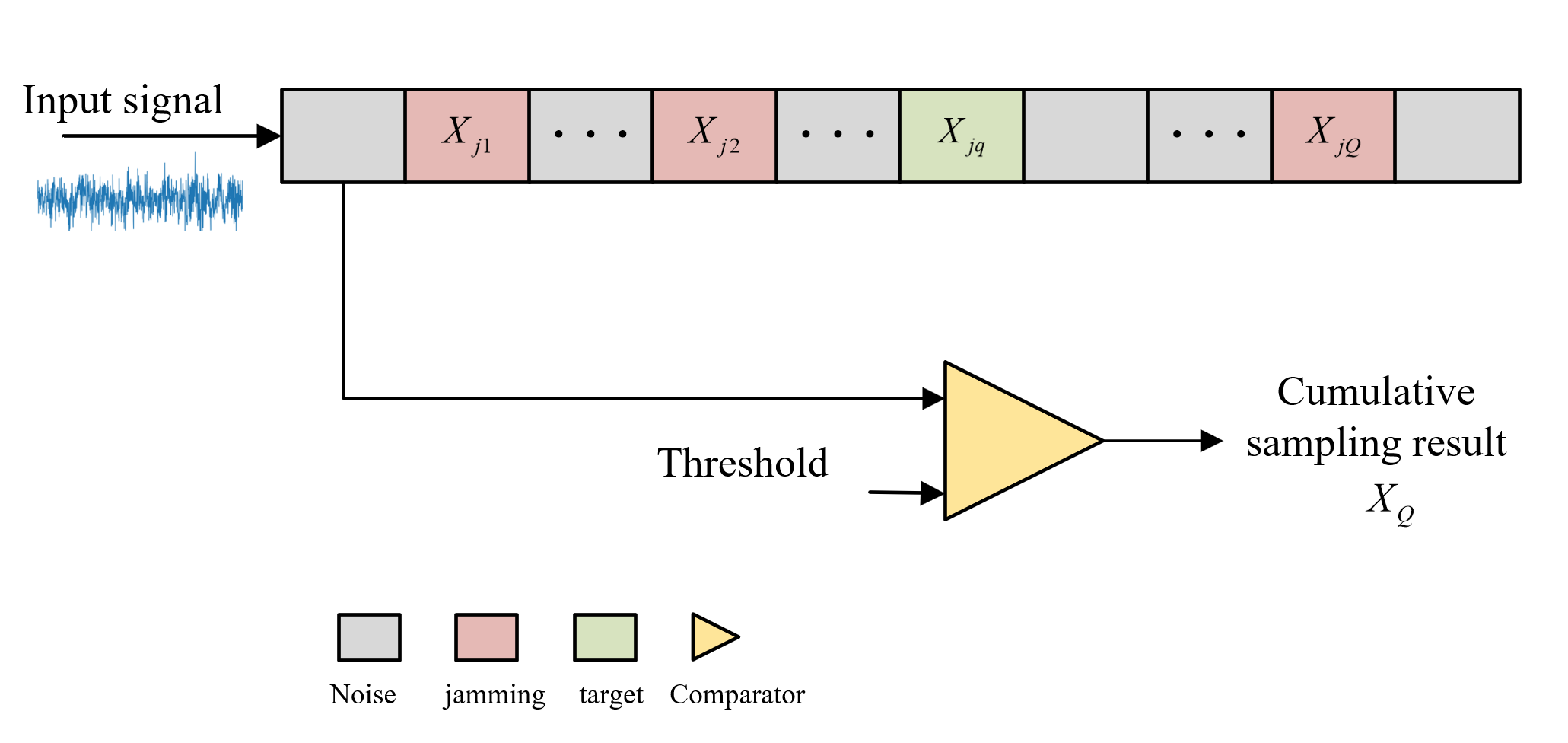}
\caption{Scheme of cumulative sampling}
\label{fig_3}
\end{figure}

\begin{equation}
{{\bf{R}}_{{\bf{X}}q}} = E\left[ {{{\bf{X}}_q}{\bf{X}}_q^H} \right] = \sum\limits_{m = 1}^\rho  {{\lambda _m}} {{\bf{u}}_m}{\bf{u}}_m^H + \sum\limits_{m = \rho  + 1}^{MN} {{\lambda _m}} {{\bf{u}}_m}{\bf{u}}_m^H\
\label{eq27}
\end{equation}

Next, sort the eigenvectors $\mathbf{u}_{m}$obtained from eigenvalue decomposition in descending order according to their eigenvalues $\lambda_{m}$. Then, calculate the sum of correlations between the top $\rho$ eigenvectors and the expected target steering vector, denoted as ${\Gamma _q}$.

\begin{equation}
{\Gamma _q} = \sum\limits_{m = 1}^\rho  {{{\left| {{\bf{u}}_m^{\rm{H}}{{\bf{v}}_{s0}}} \right|}^2}} \
\label{eq28}
\end{equation}

Then, set a threshold $\zeta$  to  determine the target , comparing the difference in correlation coefficients $\Delta \Gamma_{q-1}$ obtained from adjacent samples.

\begin{equation}
\begin{array}{c}
\Delta \Gamma_{q-1}=\left|\Gamma_{q}-\Gamma_{q-1}\right| \geq \zeta
\end{array}
\label{eq29}
\end{equation}
where $\Delta \Gamma_{q-1}$ is the difference in correlation coefficients between adjacent samples. If the difference in correlation coefficients $\Delta \Gamma_{q-1}$ is greater than $\zeta$, it can be roughly inferred that the target is located at the range gate of the $q$-th sampling.

Since the true target is located at the range bin of the $q$-th sampling, which will enhance the value of  $\Gamma_{q}$. If this sampling includes the true target, the obtained $\Gamma_{q}$ will be much larger than that obtained $\Gamma_{q-1}$ from previous cumulative samplings.

If the difference in correlation coefficients is less than the threshold, proceed the $(q+1)$th samplings, continuing the calculation according to Equation(\ref{eq27}), until all data exceeding the threshold $\eta$ have been cumulatively sampled.

If after $q=Q$ samplings the cumulatively sampled data still do not satisfy the decision of target threshold condition, it can be assumed that the sampled data not contain target signals. These sampled data can be utilized in constructing the adaptive weight vector.

A summary of the procedure of the CS-NHSS algorithm in Algorithm 1.

\begin{algorithm}[!h]
    \caption{CS-NHSS}
    \label{alg:AOA}
    \renewcommand{\algorithmicrequire}{\textbf{Input:}}
    \renewcommand{\algorithmicensure}{\textbf{Output:}}
    \begin{algorithmic}[1]
        \REQUIRE $\mathbf{x}, \mathbf{v}_{s 0}, \eta, \chi, \zeta$   
        \ENSURE the range bin of target     
        
        \STATE  Compute $\mathbf{R}_{x} $ and eigenvalue via(\ref{eq23})
        \STATE  Obtain $\gamma,\rho$ via(\ref{eq24})
        \IF {$\gamma>\chi$}
            \STATE Select the data $\mathbf{x}_{jq}(\mathbf{x}_{jq}>\eta)$;
            \FOR{each $q \in [1,Q]$}
                \STATE Construction the matrix$\mathbf{X}_{q}=\left[\mathbf{x}_{j 1}, \mathbf{x}_{j 2}, \ldots ., \mathbf{x}_{j q}\right]$;
                \STATE Perform eigenvalue decomposition on $\mathbf{X}_{q}$ via(\ref{eq27});
                \STATE Compute $\Gamma_{q}$ via(\ref{eq28});
                \STATE Compute $\Delta \Gamma_{q}$ via(\ref{eq29});
                \IF {$\Delta \Gamma_{q}>\zeta$}
                    \STATE Determine the range bin of target; break;
                \ENDIF
            \ENDFOR
            \STATE No-target decision; 
            \STATE Construct the noise subspace directly;
        \ENDIF
    \end{algorithmic}
\end{algorithm}

\subsection{Noise Subspace-based Jamming Mitigation (NSJM)}

The CS-NHSS algorithm determines the location of true target, enabling the elimination of the corresponding range gate. In this condition, we can construct the pure interference-plus-noise covariance matrix $\mathbf{R}_{j+n}$ without true target signal. Therefore, the interference-plus-noise covariance matrix $\mathbf{R}_{j+n}$ can be decomposed into the following formula:

\begin{equation}
{{\bf{R}}_{j + n}} = \sum\limits_{k = 1}^{MN} {{\lambda _k}} {{\bf{u}}_k}{\bf{u}}_k^{\rm{H}} = {{\bf{U}}_j}{\Lambda _j}{\bf{U}}_j^{\rm{H}} + {{\bf{U}}_n}{\Lambda _n}{\bf{U}}_n^{\rm{H}}
\end{equation}
where ${\Lambda _j}{\rm{ = [}}{\lambda _1}{\rm{,}}{\lambda _2}{\rm{,}} \cdot  \cdot  \cdot {\rm{,}}{\lambda _{\rho  - 1}}{\rm{]}}$ and ${\Lambda _j}{\rm{ = [}}{\lambda _\rho }{\rm{,}}{\lambda _{\rho  + 1}}{\rm{,}} \cdot  \cdot  \cdot {\rm{,}}{\lambda _{MN}}{\rm{]}}$ represents the diagonal matrix composed of its corresponding eigenvalues. Based on the orthogonal features of the jamming subspace and the noise subspace, the subspace ${{\bf{U}}_j}$  corresponding to ${\Lambda _j}$  and the subspace  ${{\bf{U}}_n}$ corresponding to ${\Lambda _n}$  are orthogonal to each other.

According to equation(\ref{eq24}), which contains $\rho$ false target jammings , the noise subspace can be represented as:

\begin{equation}
\begin{array}{c}
\mathbf{U}_{n}=\left[\mathbf{u}_{\rho}, \mathbf{u}_{\rho+1}, \cdots, \mathbf{u}_{MN}\right]
\end{array}
\end{equation}

Therefore, a weight vector about the noise subspace can be constructed for jamming mitigation, and the expression can be written as:

\begin{equation}
\begin{array}{c}
{{\bf{W}}_n} = {{\bf{U}}_n}{\bf{U}}_n^{\rm{H}}{{\bf{v}}_{s0}}
\end{array}
\end{equation}

Using this weight, the signal after jamming mitigation can be expressed as:
\begin{equation}
\begin{array}{l}
{\bf{y}} = {\bf{W}}_n^{\rm{H}}{\bf{x}} = {{\bf{U}}_n}{\bf{U}}_n^{\rm{H}}{{\bf{v}}_{s0}}\left( {\sigma _s^2{{\bf{x}}_s} + \sum\limits_{q = 1}^Q {\sigma _q^2} {{\bf{x}}_q} + {\bf{n}}} \right)\\
 = \sigma _s^2\sum\limits_{k = \rho }^{MN} {{\bf{u}}_k^{\rm{H}}} {{\bf{v}}_{s0}}{\bf{u}}_k^{\rm{H}}{{\bf{x}}_s} + 0 + \sum\limits_{k = \rho }^{MN} {{\bf{u}}_k^{\rm{H}}} {{\bf{v}}_{s0}}{\bf{u}}_k^{\rm{H}}{\bf{n}}\\
 = \sigma _s^2{\bf{W}}_n^{\rm{H}}{{\bf{x}}_s} + {\bf{W}}_n^{\rm{H}}{\bf{n}}
\end{array}
\end{equation}

The above equation shows that the output signal comprises only target and noise components, as jamming has been effectively suppressed.  This is due to the noise subspace being non-orthogonal to the target steering vector but orthogonal to the jamming steering vector.

\section{Robust Jamming mitigation via noise subspace}


In the noise subspace algorithm, we have designed $\mathbf{W}_{n}$ to suppress jamming . The essence of this method is to utilize the noise subspace to reduce the directional response of the spatial frequency domain where the false targets are located. However, in practical scenarios, this algorithm is susceptible to errors such as range quantization errors and angular deviations. In such cases, false targets may deviate from their theoretical null  positions, leading to a decrease in the performance or even failure of the jamming mitigation algorithm.

Therefore, beamforming control theory can be utilized to simultaneously control multiple regions of the beampattern, broadening the null positions. This approach seeks to design a set of robust jamming mitigation algorithms, thereby improving their effectiveness under non-ideal conditions.

According to the characteristics of MIMO-STCA, its matched filtering consists of $MN\times1$ digital channels and can be represented as:
\begin{equation}
{\bf{W}} = {{\bf{w}}_R} \otimes {{\bf{w}}_T} = {\bf{b}}(\theta ) \otimes {\bf{a}}(R,\theta ) = {\bf{b}}({f_{\rm{R}}}) \otimes {\bf{a}}({f_{\rm{T}}})\
\end{equation}
where $\otimes$ denotes the Kronecker product, and $\bf{w}_R$ and $\bf{w}_T$ are the weights for matched filtering at the receiving end and the transmitting end, respectively.

According to Equation(\ref{eq13}), it can be inferred that when the target is located at position $\left(\theta_{0},R_{0}\right)$, its corresponding transmit spatial frequency and receive spatial frequency domains can be represented as:

\begin{equation}
\begin{array}{c}
f_{\rm{T}}^{\rm{0}} = \mu \Delta t\frac{{2{R_0}}}{c} + \frac{d}{{{\lambda _0}}}\sin \left( {{\theta _0}} \right)\\
f_{\rm{R}}^{\rm{0}} = \frac{d}{{{\lambda _0}}}\sin \left( {{\theta _0}} \right)
\end{array}\
\end{equation}

Similarly, the transmit spatial frequency and receive spatial frequency domains corresponding to the $q$-th jamming can be represented as:

\begin{equation}
\begin{array}{c}
f_{\rm{T}}^q = \mu \Delta t\frac{{2{R_q}}}{c} + \frac{d}{{{\lambda _0}}}\sin \left( {{\theta _q}} \right)\\
f_{\rm{R}}^q = \frac{d}{{{\lambda _0}}}\sin \left( {{\theta _q}} \right)
\end{array}\
\end{equation}

When the target is in the mainlobe jamming situation, that is $\theta_{0}=\theta_{q}$, the receiving frequency of the true target $ f_{\mathrm{R}}^{0}$ and jamming $ f_{\mathrm{R}}^{q}$ is the same. Therefore, the true and false targets can be distinguished in the transmit frequency domain. Consequently, it can be inferred that when the target is located at position $\tilde{\mathbf{w}}_{T}$ after optimization.

By using the minimum pattern deviation design criterion, deep nulls can be generated at the jamming locations. At other locations, the original beampattern should be maintained as much as possible to minimize the beampattern deviation.

\subsection{Basic Beampattern control theory}

The minimum pattern distortion response criterion can achieve maximized output SINR. The corresponding optimal weighted vector can be represented as:

\begin{equation}
{{\bf{w}}_{opt}} = \alpha {\bf{R}}_{j + n}^{ - 1}{\bf{a}}\left( {f_{\rm{T}}^0} \right)
\label{eq39}
\end{equation}
where $\alpha$ is a normalization factor, $\mathbf{a}\left(f_{\mathrm{T}}^{s}\right)=\mathbf{a}\left(R_{0}, \theta_{0}\right)$ is the steering vector corresponding to the position in the spatial frequency domain, and $\mathbf{R}_{j+n}$ is the interference-plus-noise covariance matrix.

\begin{equation}
{{\bf{R}}_{j + n}} = \sigma _q^2{\bf{a}}(f_T^q){{\bf{a}}^{\rm{H}}}(f_T^q) + \sigma _n^2{{\bf{I}}_N}
\end{equation}

For equation(\ref{eq39}), an in-depth analysis was performed and the matrix inversion rule was applied for derivation:

\begin{equation}
\begin{array}{l}
\begin{aligned}
{{\bf{w}}_{{\rm{opt}}}} &= \alpha {\bf{R}}_{j + n}^{ - 1}{\bf{a}}\left( {f_{\rm{T}}^{\rm{0}}} \right)\\
{\rm{      }} &= \alpha {\left( {\alpha _q^2{\bf{a}}\left( {f_{\rm{T}}^q} \right){{\bf{a}}^{\rm{H}}}\left( {f_{\rm{T}}^q} \right) + \alpha _n^2{{\bf{I}}_N}} \right)^{ - 1}}{\bf{a}}\left( {f_{\rm{T}}^{\rm{0}}} \right)\\
{\rm{      }} &= \frac{\alpha }{{\alpha _n^2}}\left( {{\bf{a}}\left( {f_{\rm{T}}^{\rm{0}}} \right) - \frac{{\frac{{\alpha _q^2}}{{\alpha _n^2}}{\bf{a}}\left( {f_{\rm{T}}^q} \right){{\bf{a}}^{\rm{H}}}\left( {f_{\rm{T}}^q} \right){\bf{a}}\left( {f_{\rm{T}}^{\rm{0}}} \right)}}{{1 + \frac{{\alpha _q^2}}{{\alpha _n^2}}\left\| {{\bf{a}}\left( {f_{\rm{T}}^q} \right)} \right\|_2^2}}} \right)
\end{aligned}
\end{array}
\end{equation}

Since $\alpha / \sigma_{n}^{2}$ does not affect the output SINR of the beam formation, it can thus be neglected. Therefore, the above formula can be rewritten as:

\begin{equation}
\begin{array}{l}
\begin{aligned} \mathbf{w}_{\text {opt }} & =\mathbf{a}\left(f_{\mathrm{T}}^{0}\right)-\frac{\frac{\alpha_{q}^{2}}{\alpha_{n}^{2}} \mathbf{a}^{\mathrm{H}}\left(f_{\mathrm{T}}^{q}\right) \mathbf{a}\left(f_{\mathrm{T}}^{0}\right)}{1+\frac{\alpha_{q}^{2}}{\alpha_{n}^{2}}\left\|\mathbf{a}\left(f_{\mathrm{T}}^{q}\right)\right\|_{2}^{2}} \mathbf{a}\left(f_{\mathrm{T}}^{q}\right) \\ & =\mathbf{w}_{0}+\xi \mathbf{a}\left(f_{\mathrm{T}}^{q}\right)\end{aligned}
\label{eq42}
\end{array}
\end{equation}
where $\mathbf{w}_{0}=\mathbf{a}\left(f_{\mathrm{T}}^{s}\right)$, $\xi=\frac{\frac{\alpha_{q}^{2}}{\alpha_{n}^{2}} \mathbf{a}^{\mathrm{H}}\left(f_{\mathrm{T}}^{q}\right) \mathbf{a}\left(f_{\mathrm{T}}^{0}\right)}{1+\frac{\alpha_{q}^{2}}{\alpha_{n}^{2}}\left\|\mathbf{a}\left(f_{\mathrm{T}}^{q}\right)\right\|_{2}^{2}}$. In the above equation, the optimal weight vector $\mathbf{w}_{opt}$ is the sum of two terms. The first term $\mathbf{w}_{0}$ denotes the initial weight vector, which cannot suppress jamming on its own but serves as a basis for deriving the optimal weight vector. The second term is the product of a complex constant and the jamming steering vector $\mathbf{a}\left(f_{\mathrm{T}}^{q}\right)$.

Based on the above analysis, the beampattern response of the optimal weight vector at the spatial frequency position $f_{\mathrm{T}}^{q}$ can be written as:

\begin{equation}
\begin{array}{c}
{G_{{{\bf{w}}_{opt}}}}\left( {f_T^q} \right) = {\bf{w}}_{opt}^{\rm{H}}{\bf{a}}(f_T^q) = \left| {{G_{{{\bf{w}}_{opt}}}}\left( {f_T^q} \right)} \right|{e^{j\angle \left( {{G_{{{\bf{w}}_{opt}}}}\left( {f_T^q} \right)} \right)}}
\end{array}
\end{equation}
where $\left| {{G_{{{\bf{w}}_{opt}}}}\left( {f_T^q} \right)} \right|$ is the beampattern magnitude response, and $\angle\left| {{G_{{{\bf{w}}_{opt}}}}\left( {f_T^q} \right)} \right|$ is the beampattern phase response.

In general, when the spatial frequency domain where the mainlobe is located is $f_{T}^{0}$, the beampattern gain is maximized. Therefore, the expression for the normalized beampattern power response of the optimal weight vector is

\begin{equation}
\begin{array}{c}
{L_{{{\bf{w}}_{opt}}}}\left( {f_T^q,f_T^0} \right) = \frac{{{G_{{{\bf{w}}_{opt}}}}\left( {f_T^q} \right)}}{{{G_{{{\bf{w}}_{opt}}}}\left( {f_T^0} \right)}}{\rm{ = }}\frac{{{\bf{w}}_{opt}^{\rm{H}}{\bf{a}}(f_T^q)}}{{{\bf{w}}_{opt}^{\rm{H}}{\bf{a}}(f_T^{\rm{0}})}}
\end{array}
\end{equation}

To achieve precise array response control, the normalized magnitude and phase responses are constrained:

\begin{equation}
\begin{array}{c}
{L_{{{\bf{w}}_{opt}}}}\left( {f_T^q,f_T^0} \right) = {\rho _q}{e^{j{\varphi _q}}}\
\end{array}
\label{eq45}
\end{equation}
where $\rho_{q}$ is the normalized magnitude response at the spatial frequency $f_{T}^{q}$, and $\varphi_{q}$ is the normalized phase response at $f_{T}^{q}$.

Inspired by the form(\ref{eq42}) of the optimal weighted vector in adaptive theory, the weighted vector can be modified by adding a correction term $\xi_{k} \mathbf{a}\left(f_{T}^{q}\right)$. This can be controlled in the spatial frequency domain of point $f_{T}^{q}$ through multiple iterations:

\begin{equation}
\begin{array}{c}
\mathbf{w}_{k}=\mathbf{w}_{k-1}+\xi_{k} \mathbf{a}\left(f_{T}^{q}\right)
\end{array}
\label{eq46}
\end{equation}

Therefore, designing the parameter $\xi_{k}$ becomes extremely important. Substituting the expression for $\mathbf{w}_{k}$ from Equation(\ref{eq46}) into the normalized response expression, then the expression(\ref{eq45}) can be written as:

\begin{equation}
\begin{array}{c}
\frac{{{{\left[ {{{\bf{w}}_{k - 1}} + {\xi _k}{\bf{a}}\left( {f_T^q} \right)} \right]}^{\rm{H}}}{\bf{a}}\left( {f_T^q} \right)}}{{{{\left[ {{{\bf{w}}_{k - 1}} + {\xi _k}{\bf{a}}\left( {f_T^q} \right)} \right]}^{\rm{H}}}{\bf{a}}\left( {f_T^0} \right)}} = {\rho _q}{e^{j{\varphi _{{q_{}}}}}}
\end{array}
\label{eq47}
\end{equation}

According to the above formula can be solved the value of $\xi_{k}$:

\begin{equation}
\begin{array}{c}
{\xi _k} = \frac{{{\rho _q}{e^{ - j{\varphi _q}}}{{\bf{a}}^{\rm{H}}}(f_T^0){{\bf{w}}_{k - 1}} - {{\bf{a}}^{\rm{H}}}(f_T^q){{\bf{w}}_{k - 1}}}}{{||{\bf{a}}(f_T^q)|{|^2} - {\rho _q}{e^{ - j{\varphi _q}}}{{\bf{a}}^{\rm{H}}}(f_T^0){\bf{a}}(f_T^q)}}\
\end{array}
\label{eq48}
\end{equation}

\subsection{Multi-Region beampattern control(MRBC)}

Using the pattern control theory above, value $\rho_{q}$ are set to control the normalized effects of this point. When conducting directional pattern control, the region is divided into the mainlobe region $\Theta_{\mathrm{m}}$ and the null region $\Theta_{\mathrm{q}}$. Assuming the normalized magnitude responses in mainlobe region and null region are $\rho_{m}$ and $\rho_{q}$ respectively.

Therefore, during the $k$-th iteration, the point that deviates the most from the constraint in the previous iteration is selected for control. This can be represented as:

\begin{equation}
\begin{array}{c}
f_{\mathrm{T}}^{q, k}=\left\{\begin{array}{ll}\arg \max _{f_{\mathrm{T}} \in \Theta_{\mathrm{m}}}\left|L_{k-1}\left(f_{\mathrm{T}}^{q, k-1}, f_{\mathrm{T}}^{0}\right)-\rho_{m}\right|, \\  
f_{\mathrm{T}} \in \Theta_{\mathrm{m}}, q=0 \\ \arg \max _{f_{\mathrm{T}} \in \Theta_{q}}\left(L_{k-1}\left(f_{\mathrm{T}}^{q, k-1}, f_{\mathrm{T}}^{0}\right)-\rho_{q}\right), \\ 
f_{\mathrm{T}} \in \Theta_{q}, q=1,2 \ldots, Q\end{array}\right.
\end{array}
\label{eq49}
\end{equation}

Therefore, multi-region control can be simplified to controlling multiple points that deviate the most. Meanwhile, during multi-region control, $\mathbf{w}_{k}$  can be rewritten as:

\begin{equation}
\begin{array}{c}
{{\bf{w}}_k} = {{\bf{w}}_{k - 1}} + \sum\limits_{q = {\rm{0}}}^Q {{\xi _{k,q}}} {\bf{a}}(f_T^{q,k})
\end{array}
\end{equation}

It should be noted that when $q=0$ corresponds to controlling the mainlobe region $f_{\mathrm{T}} \in \Theta_{\mathrm{m}}$, and $\rho_{m}=\rho_{0}$; when $q=1,2 \ldots, Q$, it corresponds to controlling the null region $f_{\mathrm{T}} \in \Theta_{\mathrm{q}}$.

When controlling a specific region, the beampattern in other regions may become distorted. Therefore, to minimize distortion in other regions according to the Minimum Pattern Distortion (MPD) criterion, the constraint condition can be written as:

\begin{equation}
\begin{array}{*{20}{l}}
{{{\min }_{{\varphi _{q,k}}}}}{\left| {{L_{{{\bf{w}}_k}}}\left( {f_T^k,f_T^0} \right) - {L_{{{\bf{w}}_{k - 1}}}}\left( {f_T^{k - 1},f_T^0} \right)} \right|{\rm{,  }}f_T^{} \notin {\rm{ }}{\Theta _m},{\Theta _q}}\\
{{\rm{ s}}{\rm{.t}}{\rm{. }}}{\left\{ \begin{array}{l}
{L_{{{\bf{w}}_k}}}\left( {f_T^{q,k}} \right) = {\rho _{k,q}}{e^{j{\varphi _{q,k}}}}\\
{{\bf{w}}_k} = {{\bf{w}}_{k - 1}} + \sum\limits_{q = 1}^Q {{\xi _{q,k}}} {\bf{a}}(f_T^{q,k})
\end{array} \right.{\rm{ }}q = 0,1,2 \cdots ,Q}\\
{}&{}
\end{array}
\end{equation}

Furthermore, according to Equation(\ref{eq48}), the parameter $\xi_{q, k}$ of the $q$-th region in the $k$-th iteration can be rewritten as:

\begin{equation}
\begin{array}{c}
\xi_{q, k}=\frac{\rho_{q, k} e^{-j \varphi_{q, k}} \mathbf{a}^{\mathrm{H}}\left(f_{T}^{0}\right) \mathbf{w}_{k-1}-\mathbf{a}^{\mathrm{H}}\left(f_{T}^{q, k}\right) \mathbf{w}_{k-1}}{\left\|\mathbf{a}\left(f_{T}^{q, k}\right)\right\|^{2}-\rho_{q, k} e^{-j \varphi_{q, k}} \mathbf{a}^{\mathrm{H}}\left(f_{T}^{0}\right) \mathbf{a}\left(f_{T}^{q, k}\right)}
\end{array}
\end{equation}

Therefore, we can solve the key parameter:
\begin{equation}
\begin{array}{c}
\varphi_{q, k}=\angle\left(\frac{\mathbf{w}_{k-1}^{\mathrm{H}} \mathbf{a}\left(f_{T}^{q, k}\right)}{\mathbf{w}_{k-1}^{\mathrm{H}} \mathbf{a}\left(f_{T}^{0}\right)}\right), q=0,1,2, \cdots, Q
\end{array}
\label{eq53}
\end{equation}

When single-point control from Equation(\ref{eq47}) is extended to multi-point control, there are $Q+1$ constraints in total. The above equation can be equivalently written as:

\begin{equation}
\left\{ \begin{array}{l}
\frac{{{{\bf{a}}^{\rm{H}}}(f_T^{0,k}){\bf{w}}_{k - 1}^{} + {{\bf{a}}^{\rm{H}}}(f_T^{0,k})\sum\limits_{q = 0}^Q {\xi _{q,k}^{}} {\bf{a}}(f_T^{q,k})}}{{{{\bf{a}}^{\rm{H}}}(f_T^0){\bf{w}}_{k - 1}^{} + {{\bf{a}}^{\rm{H}}}(f_T^0)\sum\limits_{q = 0}^Q {\xi _{q,k}^{}} {\bf{a}}(f_T^{q,k})}} = {\rho _{0,k}}{e^{j{\varphi _{0,k}}}}\\
\frac{{{{\bf{a}}^{\rm{H}}}(f_T^{1,k}){\bf{w}}_{k - 1}^{} + {{\bf{a}}^{\rm{H}}}(f_T^{1,k})\sum\limits_{q = 0}^Q {\xi _{q,k}^{}} {\bf{a}}(f_T^{q,k})}}{{{{\bf{a}}^{\rm{H}}}(f_T^0){\bf{w}}_{k - 1}^{} + {{\bf{a}}^{\rm{H}}}(f_T^0)\sum\limits_{q = 0}^Q {\xi _{q,k}^{}} {\bf{a}}(f_T^{q,k})}} = {\rho _{1,k}}{e^{j{\varphi _{1,k}}}}\\
{\rm{                                }} \vdots \\
\frac{{{{\bf{a}}^{\rm{H}}}(f_T^{Q,k}){\bf{w}}_{k - 1}^{} + {{\bf{a}}^{\rm{H}}}(f_T^{Q,k})\sum\limits_{q = 0}^Q {\xi _{q,k}^{}} {\bf{a}}(f_T^{q,k})}}{{{{\bf{a}}^{\rm{H}}}(f_T^0){\bf{w}}_{k - 1}^{} + {{\bf{a}}^{\rm{H}}}(f_T^0)\sum\limits_{q = 0}^Q {\xi _{q,k}^{}} {\bf{a}}(f_T^{q,k})}} = {\rho _{Q,k}}{e^{j{\varphi _{1,k}}}}
\end{array} \right.
\label{eq54}
\end{equation}

Thus, the above formula can be written as a matrix, which can be expressed as:

\begin{equation}
\begin{array}{c}
\frac{\mathbf{A}_{k}^{\mathrm{H}} \mathbf{w}_{k-1}+\mathbf{A}_{k}^{\mathrm{H}} \mathbf{A}_{k} \Xi_{k}}{\left(\mathbf{a}^{\mathrm{H}}\left(f_{T}^{0}\right) \mathbf{w}_{k-1}+\mathbf{a}^{\mathrm{H}}\left(f_{T}^{0}\right) \mathbf{A}_{k} \Xi_{k}\right)}=\boldsymbol{\Psi}_{k}
\end{array}
\end{equation}
where

\begin{equation}
\begin{array}{c}
{{\bf{A}}_k} = \left[ {{\bf{a}}(f_T^{0,k}),{\bf{a}}(f_T^{1,k}), \cdots ,{\bf{a}}(f_T^{Q,k})} \right]
\end{array}
\label{eq56}
\end{equation}

\begin{equation}
\begin{array}{c}
{{\bf{\Psi }}_k} = {\left[ {{\rho _{0,k}}{e^{ - j{\varphi _{0,k}}}},{\rho _{1,k}}{e^{ - j{\varphi _{1,k}}}}, \cdots ,{\rho _{Q,k}}{e^{ - j{\varphi _{Q,K}}}}} \right]^{\rm{T}}}
\end{array}
\label{eq57}
\end{equation}

Hence, The matrix composed of the key parameter $\xi_{q, k}$ can be expressed as:

\begin{equation}
\begin{array}{c}
\begin{aligned} \Xi_{k} & =\left[\xi_{0, k}, \xi_{1, k}, \cdots, \xi_{q, k}\right]^{\mathrm{T}} \\ & =\frac{\left(\mathbf{a}^{\mathrm{H}}\left(f_{T}^{0}\right) \mathbf{w}_{k-1} \boldsymbol{\Psi}_{k}-\mathbf{A}_{k}^{\mathrm{H}} \mathbf{w}_{k-1}\right)}{\left(\mathbf{A}_{k}^{\mathrm{H}} \mathbf{A}_{k}-\mathbf{\Psi}_{k} \mathbf{a}\left(f_{T}^{0}\right) \mathbf{A}_{k}\right)}\end{aligned}
\end{array}
\label{eq58}
\end{equation}

Thus, multipoint control written in matrix form can be rewritten as:

\begin{equation}
\begin{array}{c}
\mathbf{w}_{k}=\mathbf{w}_{k-1}+\mathbf{A}_{k} \Xi_{k}
\end{array}
\end{equation}

Furthermore, according to Equation(\ref{eq58}),the critical parameters $\varphi_{q, k}$ can be obtained from Equation(\ref{eq53}).

\subsection{The amplitude constraint of the synthesized beampattern}

In the previous section, the weights $\mathbf{w}_{k}$ are controlled iteratively. Therefore, it is necessary to set stopping condition to terminate the iteration when the amplitude constraints are satisfied.

\begin{equation}
\begin{array}{c}
\varepsilon_{q, k}=\left\{\begin{array}{ll}\max \left|L_{k}\left(f_{\mathrm{T}}^{0, k}, f_{\mathrm{T}}^{\mathrm{s}}\right)-\rho_{\mathrm{m}}\right|, & f_{\mathrm{T}}^{0, k} \in \Theta_{\mathrm{m}}, q=0 \\ \max \left(L_{k}\left(f_{\mathrm{T}}^{q, k}, f_{\mathrm{T}}^{\mathrm{s}}\right)-\rho_{\mathrm{n}}\right), & f_{\mathrm{T}}^{q, k} \in \Theta_{q}, {q \in [1,Q]}\end{array}\right.
\end{array}
\end{equation}

The above equation provides the calculation method for the maximum deviation between the $q$-th region in the $k$-th iteration and the ideal beampattern.

Then, set the threshold for the maximum deviation in the mainlobe region as $\vartheta_{m}$ and for the null region as $\vartheta_{n}$. If $\varepsilon_{q, k}$ satisfies the following conditions, the beampattern is considered to satisfy the design requirements after the $k$-th iteration.

\begin{equation}
\begin{array}{c}
\left\{\begin{array}{c}\varepsilon_{q, k} \leq \vartheta_{m}, \mathrm{q}=0 \\ \varepsilon_{q, k} \leq \vartheta_{n}, q=1,2 \ldots, Q\end{array}\right.
\end{array}
\label{eq61}
\end{equation}

\subsection{Robust jamming mitigation weight}

According to the result of the iteration, make the weight $\tilde{\mathbf{w}}_{T}=\mathbf{w}_{k}$.

There, the Robust jamming suppression weights can be expressed as:

\begin{equation}
\begin{array}{c}
\tilde{\mathbf{W}}=\mathbf{w}_{R} \otimes \tilde{\mathbf{w}}_{T}
\end{array}
\end{equation}

Then, the desired target steering vector $ {{\bf{v}}_{s0}}$ in ${{\bf{U}}_n}$ can be replaced by $\tilde{\mathbf{W}}$, and a new set of robust jamming suppression weight ${{\bf{W}}_s}$ can be obtained by bringing it into the noise subspace.

\begin{equation}
\begin{array}{c}
{{\bf{W}}_s}{\rm{  = }}{{\bf{U}}_n}{\bf{U}}_n^{\rm{H}}{\bf{\tilde W}}
\end{array}
\label{eq63}
\end{equation}

In general, the above algorithms can be summarized in the following pseudocode form

\begin{algorithm}[!h]
    \caption{ Robust Jamming mitigation via noise subspace(RJNS)}
    \label{alg:AOA}
    \renewcommand{\algorithmicrequire}{\textbf{Input:}}
    \renewcommand{\algorithmicensure}{\textbf{Output:}}
    \begin{algorithmic}[1]
        \REQUIRE $\mathbf{a}\left(f_{\mathrm{T}}^{0}\right), \rho_{q}, \Theta_{m}, \Theta_{q}, \vartheta_{m}, \vartheta_{n}$  
        \ENSURE ${{\bf{W}}_s}$   
        
        \REPEAT 
        \STATE k = k+1;
        \STATE Select multiple control point $f_{\mathrm{T}}^{q, k-1}$ in region via(\ref{eq49});
        \STATE Set normalized magnitude via(\ref{eq45});
        \STATE Compute phase parameter via(\ref{eq53});
        \STATE Construct parameter matrix via(\ref{eq56})(\ref{eq57}),\\and compute $\Xi_{k}(\ref{eq58})$;
        \STATE Update $\mathbf{w}_{k}=\mathbf{w}_{k-1}+\mathbf{A}_{k} \Xi_{k}$;
        
        \UNTIL{$\varepsilon_{q, k}$ is satisfied condition(\ref{eq61});} 
        \STATE $\tilde{\mathbf{w}}_{T}=\mathbf{w}_{k}$;
        \STATE Construct anti-inference weight via(\ref{eq63});
    \end{algorithmic}
\end{algorithm}

\section{Simulation experiment}

In this section, simulation experiments are conducted to verify the effectiveness of the proposed methods in target localization and jamming suppression algorithms under mainlobe false target jamming conditions. The simulation parameters are shown in the following Table I and Table II, which are MIMO-STCA Radar parameter table and target parameter table respectively.

\begin{table}[!h]
\caption{Parameter of MIMO-STCA Radar \label{tab:table1}}
\centering
\begin{tabular}{llll}
			\hline
            \hline
			\textbf{Parameter} & \textbf{Value} & \textbf{Parameter}  & \textbf{Value} \\ \hline                 
            M & 16 & N & 16 \\ 
            Carrier frequency & 10GHZ & Sampling frequency & 10MHz \\ 
            PRF & 5KHz & Bandwidth & 10MHz  \\ 
            Range gate number & 2000 & Pulse number & 30 \\ 
            Transmit delay$\tau$ & 0.1022us & Pulse time & 1us \\ \hline
		\end{tabular}
\end{table}

\begin{table}[!h]
\caption{Parameter of Target\label{tab:table2}}
\centering
\begin{tabular}{lllll}
			\hline
            \hline
			\textbf{ } & \textbf{True } & \textbf{False 1}  & \textbf{False 2} & \textbf{False 3} \\ \hline           
            angle(\degree) & 0 & 0 & 0 & 0 \\ 
            range(km) & 43 & 64 & 66 & 84 \\ 
            SNR/JNR(dB) & 20 & 30 & 30 & 30  \\ 
            Range bin & 1221 & 321 & 431 & 1601 \\ \hline
		\end{tabular}
\end{table}
\subsection{Target rough positioning}

For the noise subspace jamming mitigation (NSJM) algorithm designed above, a pure interference-plus-noise covariance matrix needs to be obtained. First, it is necessary to locate the true target range bin information, and then set $\eta=7(\mathrm{~dB})$ the sampling threshold for sampling:
\begin{figure}[!h]
\centering
\includegraphics[width=2.6in]{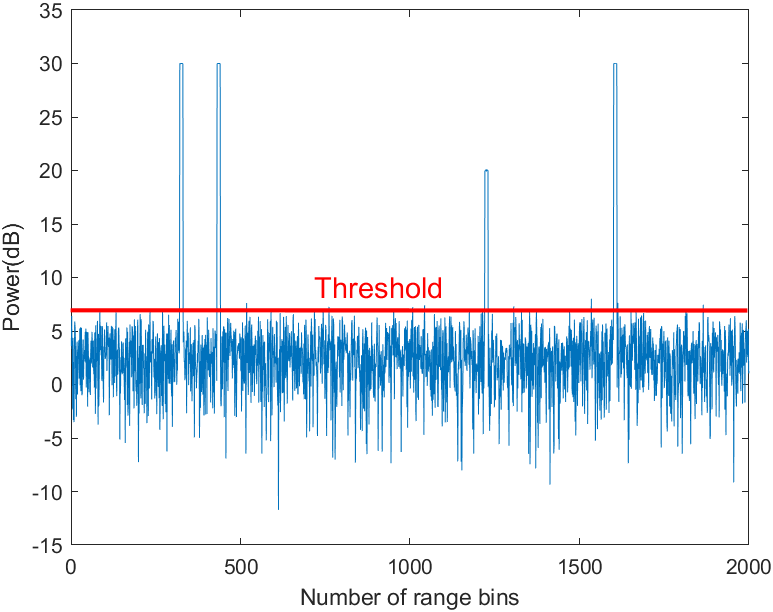}
\caption{ multi-mainlobe false target jamming.}
\label{fig_4}
\end{figure}

The echo signals are fully sampled and the covariance matrix is constructed. Decomposition yields eigenvalues and eigenvectors. The eigenvalues are sorted in descending order as shown in Fig. 5(a), and the correlation coefficients of the eigenvectors corresponding to each eigenvalue with the desired target steering vector are shown. It can be seen that under this simulation parameter, there is a target and 3 false target jammings. A correlation threshold of $\chi {\rm{ = 150}}$ is set. Therefore, signals greater than the threshold $\eta$ will be used in cumulative sampling target localization algorithm.

\begin{figure}[!h]
\centering
\subfloat[Eigenvalue]{
		\includegraphics[width=2.6in]{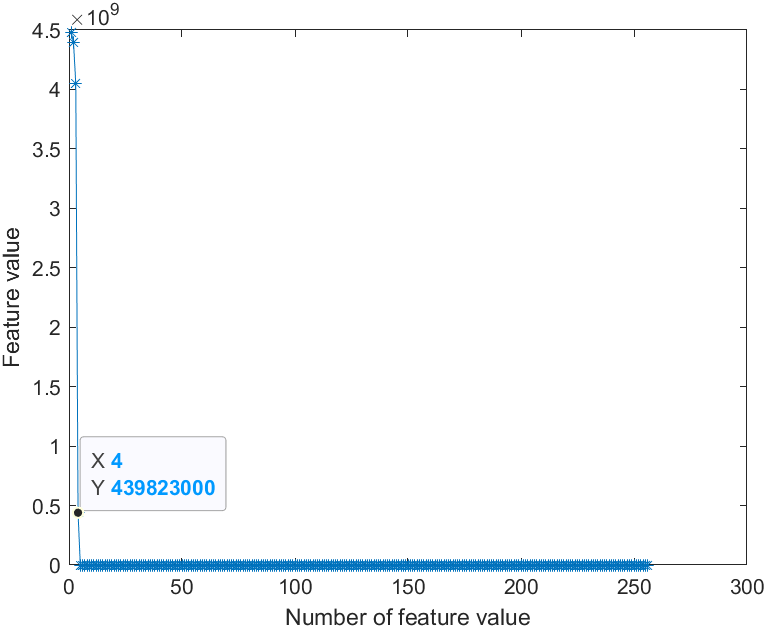}}\\
\subfloat[Correlation coefficient]{
		\includegraphics[width=2.6in]{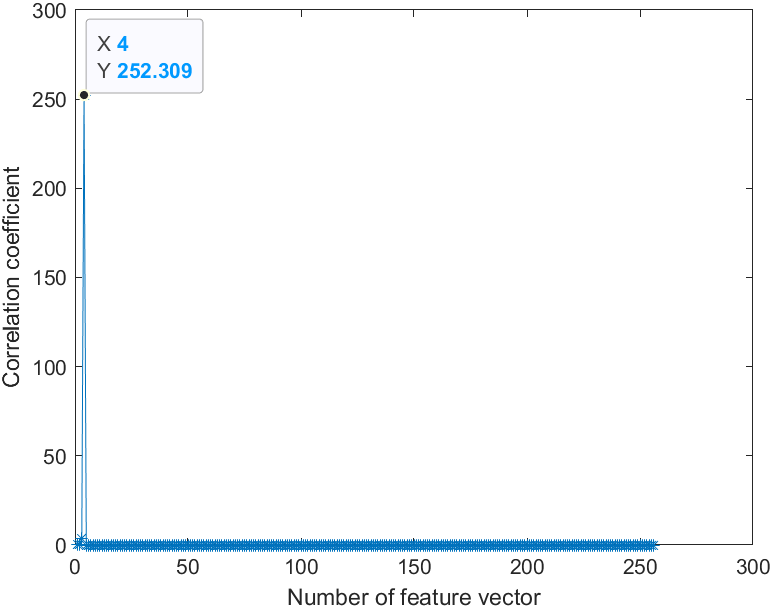}}
\caption{eigenvalue decomposition}
\label{fig_5}
\end{figure}

Fig. 6 shows the correlation $\Delta \Gamma$ decision. By setting a correlation threshold $\zeta=125$, when the correlation $\Delta \Gamma_{q}$ is greater than $\zeta$ after the $q$-th sampling, the target is determined to be in the $q$-th sampling interval. From Fig. 6, when $q=3$, $\Delta \Gamma_{q} >\zeta=125$ , so the true target is determined to be in the 3rd sampling interval, corresponding to the 1221st range bin, which is consistent with the simulated position of the target.
\begin{figure}[!h]
\centering
\includegraphics[width=3in]{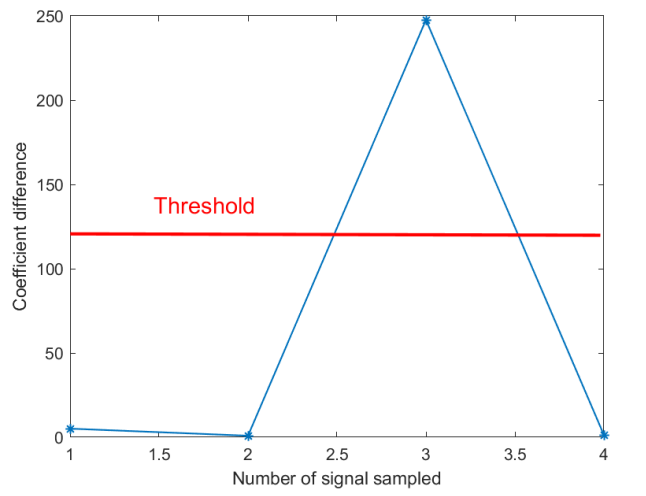}
\caption{Cumulative sampling threshold decision}
\label{fig_6}
\end{figure}

\subsection{Jamming mitigation  algorithm based on noise subspace}

By simulation 1, the range bin where the true target is located is obtained. With this information, the signal corresponding to the true target is removed to construct a pure interference-plus-noise covariance matrix, based on which the weight vectors are built using the noise subspace. Fig, 7 shows the pulse compression results after jamming suppression using the noise subspace algorithm. It can be observed that this method effectively suppresses mainlobe jamming.
\begin{figure}[!h]
\centering
\includegraphics[width=3in]{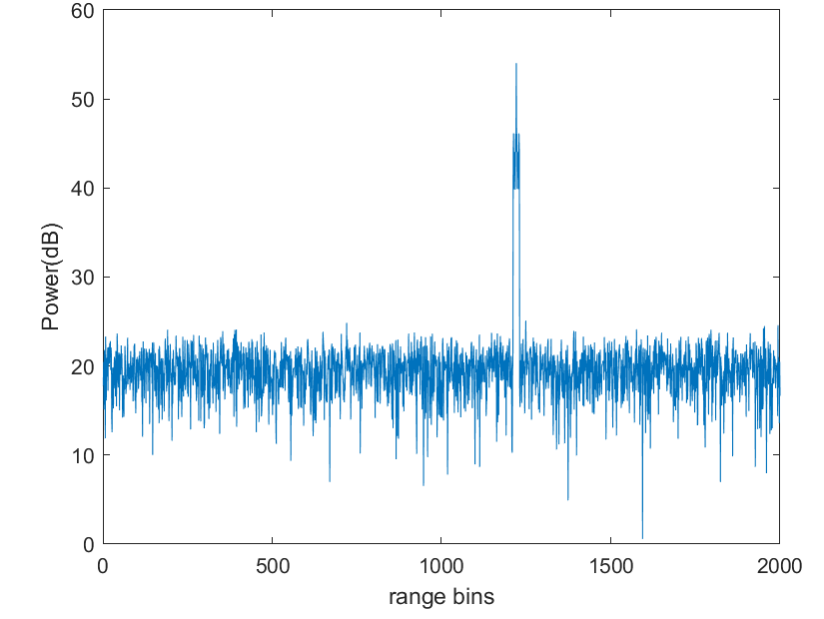}
\caption{NSJS algorithm result}
\label{fig_3}
\end{figure}

Fig. 8(a) validates that STCA-MIMO can distinguish between true and false targets in the transmit-receive Capon spectrum. Fig. 8(b) shows the transmit-receive Capon spectrum constructed by the weight vectors from the noise subspace. By forming a mainlobe at the target position and nulls in the jamming region for jamming suppression, false targets are suppressed in the transmit spatial frequency domain.

\begin{figure}[!h]
\centering
\subfloat[true and false targets in Capon spectrum]{
		\includegraphics[width=3in]{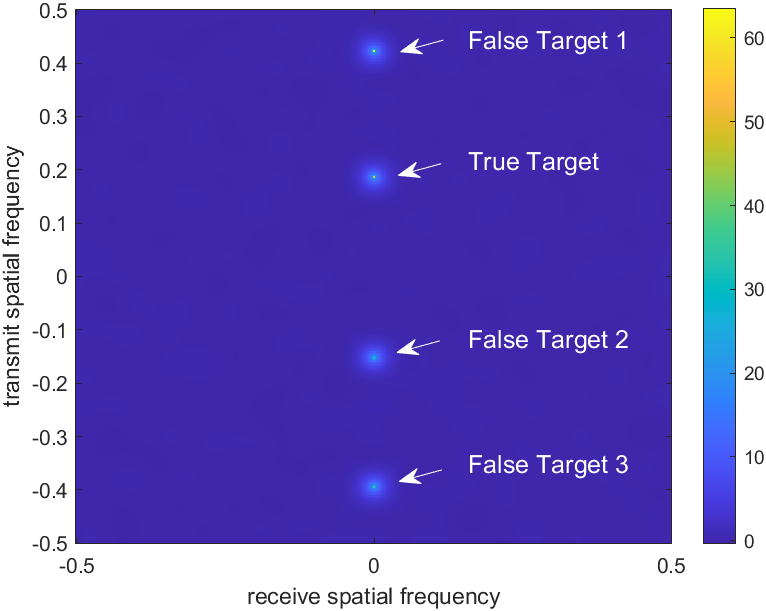}}\\
\subfloat[beampattern of NSJS]{
		\includegraphics[width=3in]{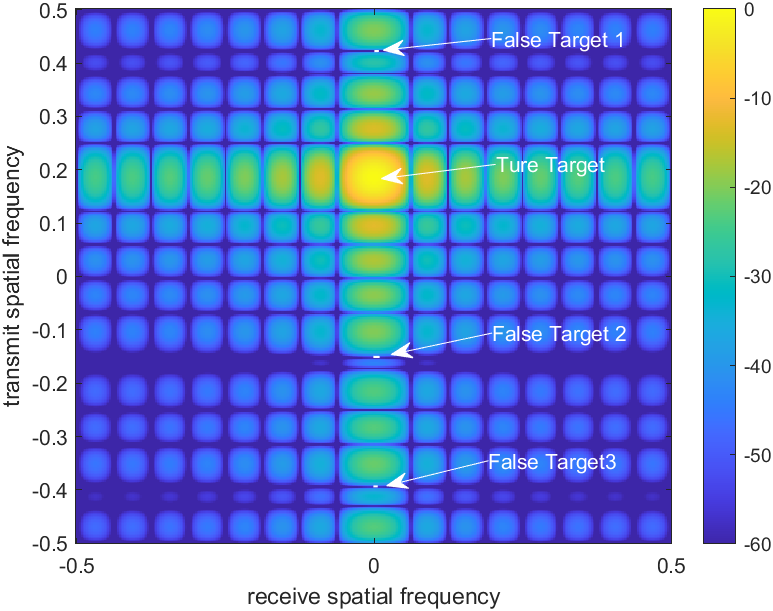}}
\caption{Adaptive jamming suppression}
\label{fig_8}
\end{figure}

Similarly, as shown in Fig. 9, jamming mitigation can be achieved when the true target is located in the 651st range bin, and the false targets are located in the 501st and 1201st range bins, respectively.

\begin{figure*}[!h]
\centering
\subfloat[ 	Capon spectrum ]{
		\includegraphics[width=3.1in]{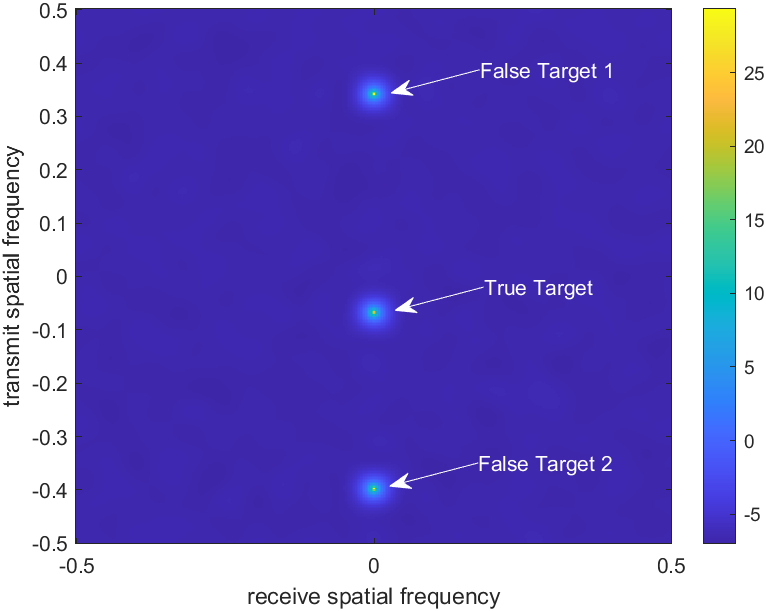}}
\subfloat[beampattern]{
		\includegraphics[width=3.1in]{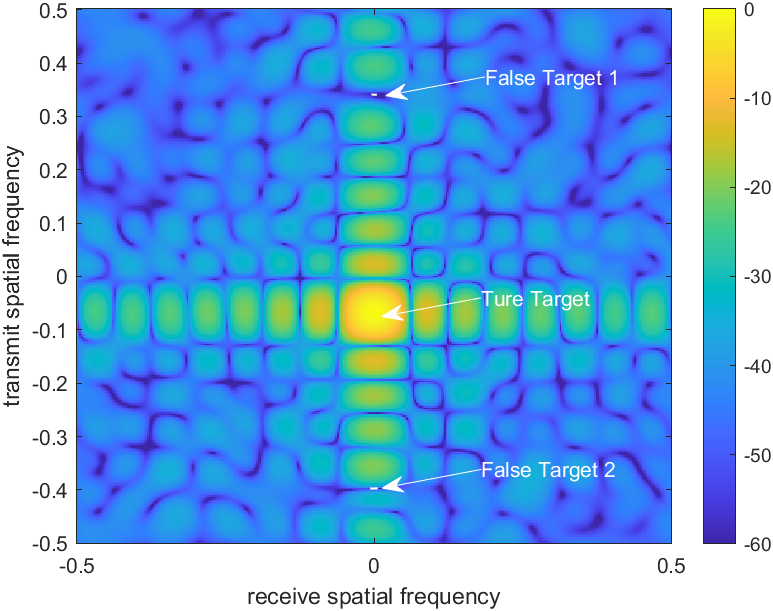}}\\
\subfloat[sectional view of transmit spatial frequency]{
		\includegraphics[width=3.1in]{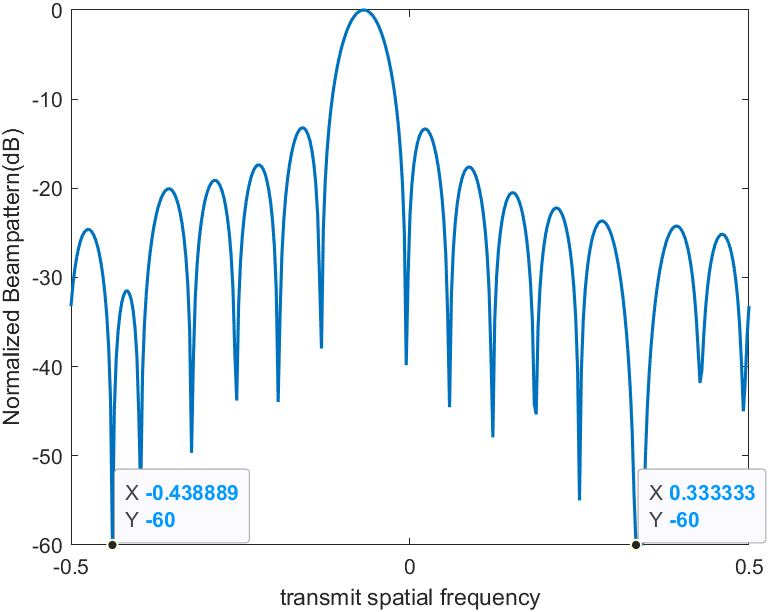}}
\caption{two false target jamming scenario}
\label{fig_9}
\end{figure*}

Therefore, the NSRJ algorithm can suppress $M-1$ mainlobe Jammings at most.

\subsection{Robust jamming mitigation algorithm}

\begin{figure*}[!ht]
\centering
\subfloat[ DOA errors ]{
		\includegraphics[width=3.5in]{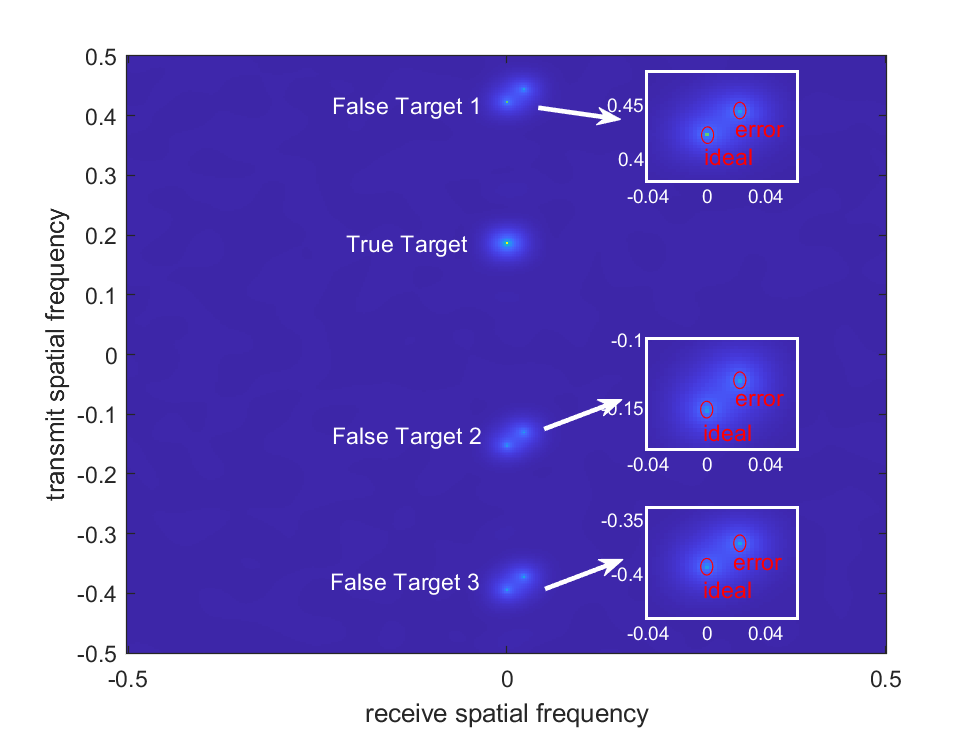}}
\subfloat[Range quantization error]{
		\includegraphics[width=3.5in]{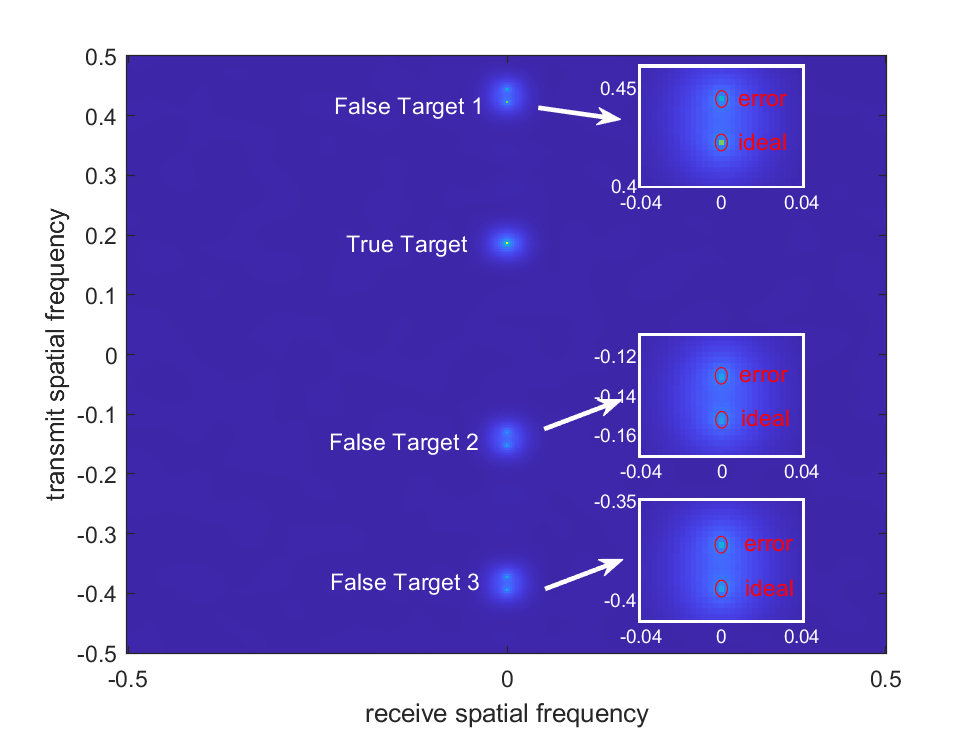}}

\caption{Effect of DOA and Range quantization error}
\label{fig_10}
\end{figure*}

Fig. 10 illustrates the shift in the interference position within the Capon spectrum caused by a 1 degree DOA error and a 1 range bins quantization error for false targets. Therefore, robust interference suppression is necessary.

\begin{figure*}[!ht]
\centering
\subfloat[ multi-null regions ]{
		\includegraphics[scale=0.6]{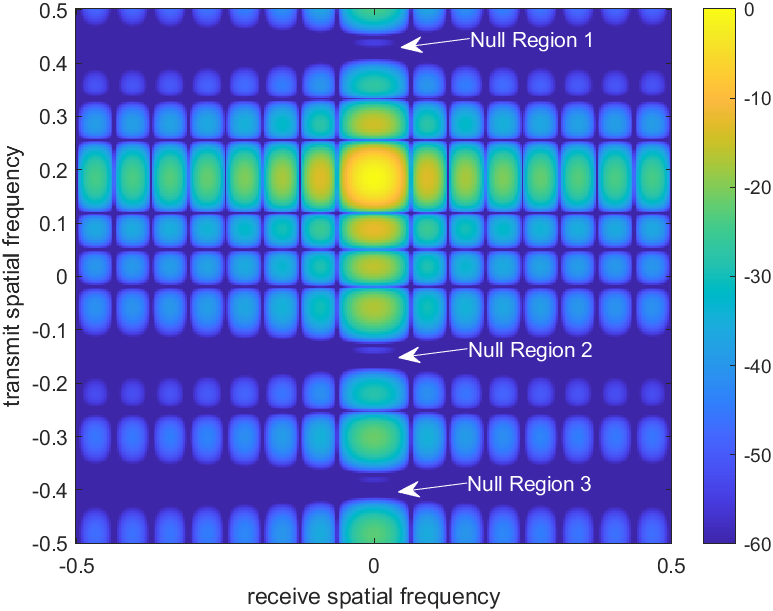}}
\subfloat[transmit spatial pattern]{
		\includegraphics[scale=0.6]{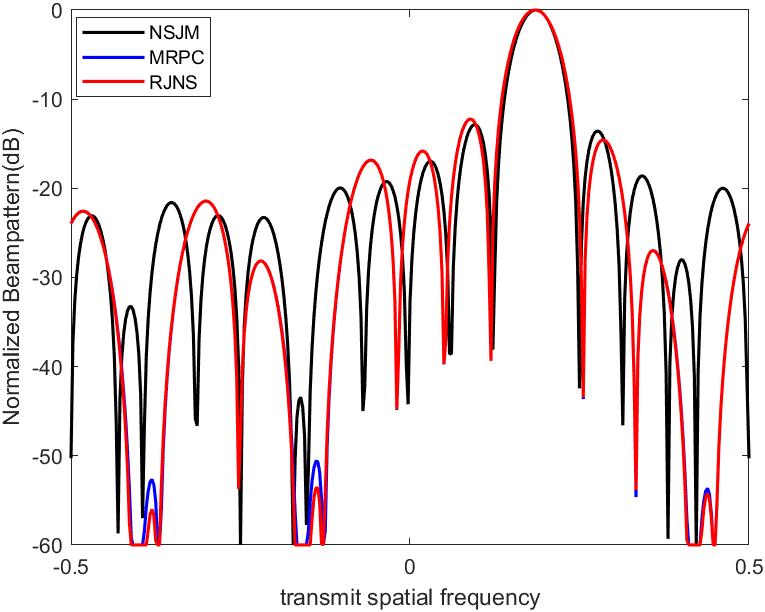}}
\\
\subfloat[mainlobe and multi-null regions]{
		\includegraphics[scale=0.6]{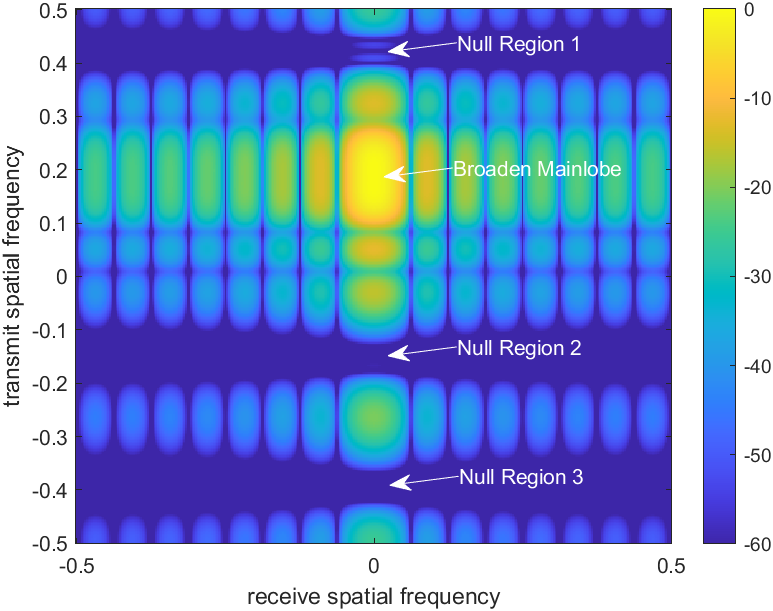}}
\subfloat[ transmit spatial pattern]{
		\includegraphics[scale=0.6]{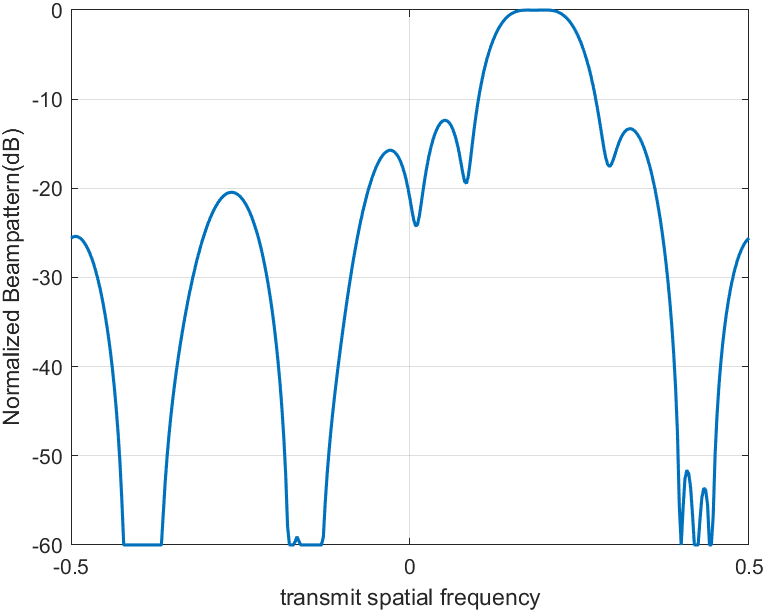}}
\caption{Robust Beampattern control}
\label{fig_10}
\end{figure*}

Fig. 11(a) shows the two-dimensional transmit spatial pattern after robust jamming suppression weighting. As depicted in the figure, this algorithm can form deeper and wider nulls at jamming positions. Even in the presence of measurement errors, it can maintain good jamming suppression performance. Fig. 11(b) illustrates the cross-section of the transmit spatial frequency domain when the receive spatial frequency domain is zero. As shown, this algorithm can precisely control the null positions of multiple mainlobe jammings at -0.15±0.02, -0.39±0.02, and 0.42±0.02 (where the spatial frequency ±0.02 corresponds to an angle deviation of 1.2°), forming deep nulls at -50dB. Compared to the NSRJ algorithm, it extends the depth and width of nulls in this region. Compared to the MRBC algorithm, it has deeper null regions.

Fig. 11(c) presents the robust anti-mainlobe jamming beamforming pattern obtained by the proposed algorithm, while Fig. 11(d) depicts the profile when the receive spatial frequency is zero. It is evident that a wide null at -50dB is formed at the false target position, while a flat-top main lobe is formed at the true target position. This effectively overcomes the problem of decreased jamming suppression performance caused by errors in the presence of true and false targets.

\subsection{Jamming mitigation  performance analysis}

In the assessment of jamming mitigation  performance, the SINR metric is commonly used for quantitative evaluation. The expression of SINR as follow:

\begin{equation}
\begin{array}{c}\operatorname{SINR}=\frac{\sigma_{s}^{2}\left|\mathbf{w}^{H} \mathbf{a}\left(R_{0}, \theta_{0}\right)\right|^{2}}{\mathbf{w}^{H} \mathbf{R}_{j+ n} \mathbf{w}}
\end{array}
\end{equation}

But in practical engineering, it is difficult to obtain a pure  interference-plus-noise covariance matrix. The covariance matrix $\mathbf{R}_{X}=E\left\{\mathbf{x x}^{H}\right\}$ of the received signals can be used as a substitute. The following Fig. 12 presents the SINR simulation results of the proposed algorithm in this paper.

\begin{figure}[!h]
\centering
\includegraphics[width=3.5in]{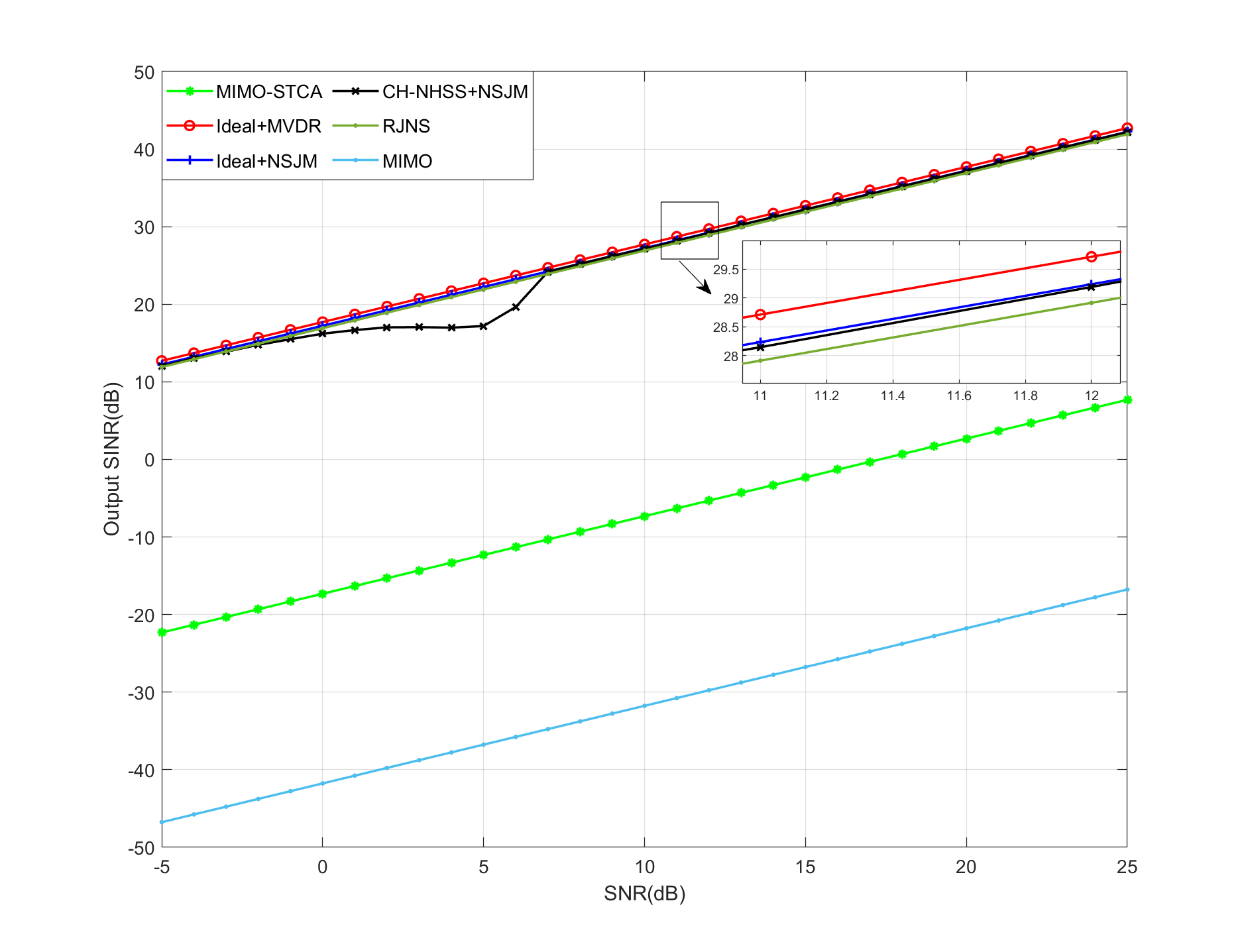}
\caption{Jamming mitigation performance}
\label{fig_3}
\end{figure}

According to the Fig.12, MIMO-STCA significantly improves interference performance under mainlobe jamming by approximately 25dB compared to traditional MIMO.

Under ideal conditions, the proposed NSJM algorithm performs similarly to the MVDR algorithm.

Meanwhile, the proposed CS-NHSS combined with the NSJM algorithm may fail to detect the target at low SNR. In such cases, jamming suppression is carried out directly using the covariance matrix to form the noise subspace. However, as the SNR increases, directly obtaining the noise subspace from the correlation matrix will lead to performance degradation.

However, when the SNR exceeds a certain threshold does the CS-NHSS target coarse positioning algorithm become effective, allowing for the acquisition of a pure interference-plus-noise covariance matrix. At this point, the jamming suppression performance approaches the ideal scenario.

Furthermore, the proposed RJNS does not affect the performance of SINR. On this basis, it can broaden the nulls region, thereby enhancing the robustness of jamming suppression.

\section{Conclusion}

In this paper, we has proposed two methods to suppress mainlobe jamming by MIMO-STCA. 

Firstly, the NSJM method has been proposed by utilizing the orthogonal properties between the noise subspace and jamming subspace to achieve jamming suppression. In this aspect, the CS-NHSS method has been proposed to roughly locate the target, so as to obtain the pure interference-plus-noise covariance matrix and pure noise subspace.

The RJNS method has been proposed to achieve robust jamming mitigation in scenarios where both target jamming and false target jamming are affected by DoA or range quantization errors. Building on the NSJM algorithm, RJNS incorporated the adaptive theory of beampattern control to achieve a flat-top mainlobe and wider nulls. This enhances the robustness of MIMO-STCA radar systems, enabling them to maintain effective mitigation of mainlobe jamming even in the presence of errors.

Future research will focus on extending the proposed method to handle scenarios with multiple targets as well as those affected by clutter interference.

\begin{IEEEbiography}[{\includegraphics[width=1in,height=1.25in,clip,keepaspectratio]{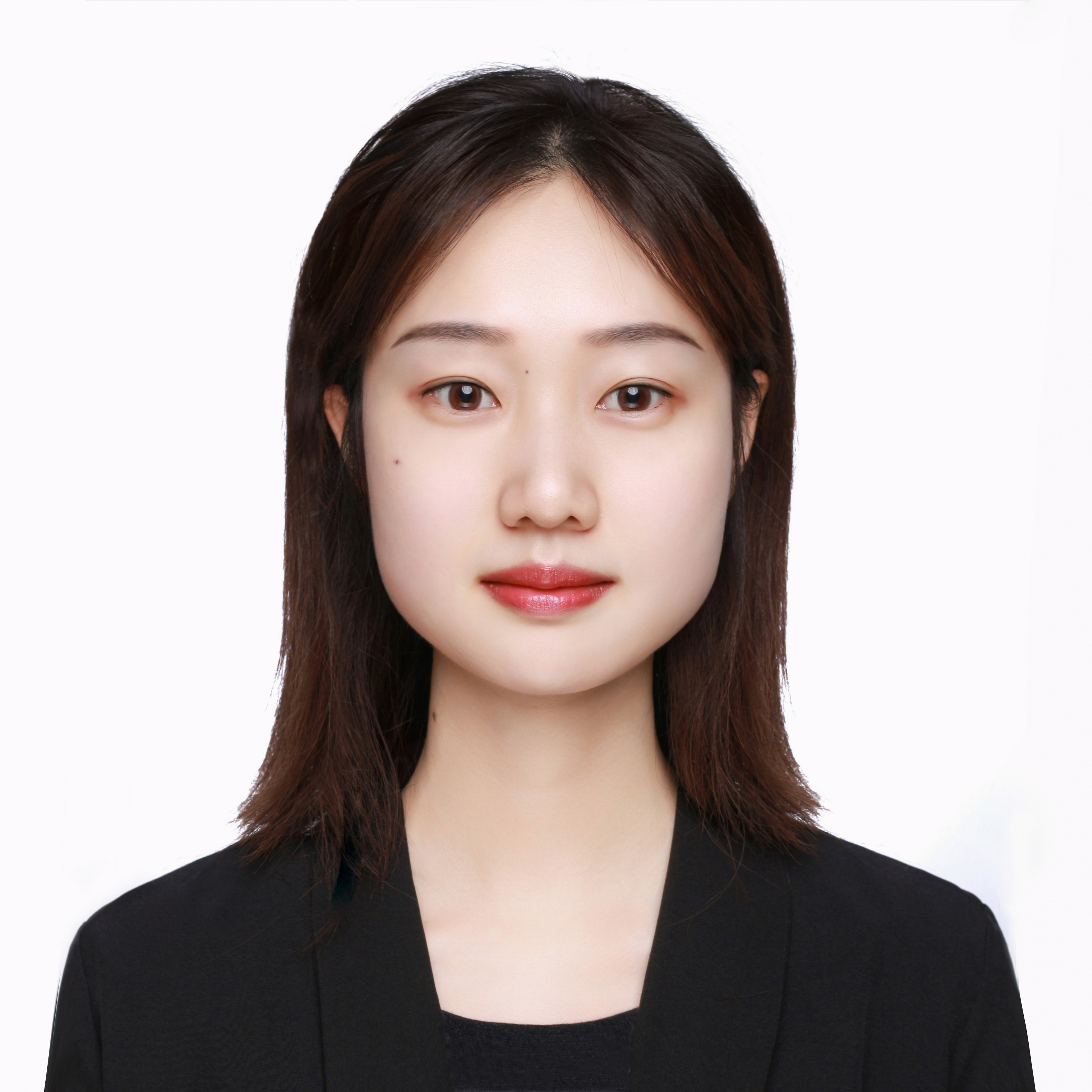}}]{Huake Wang}
received the B.S. degree in electronics engineering, and the Ph.D. degree in signal and information processing from Xidian University, Xi'an, China, in 2015 and 2020, respectively. She was a Visiting Ph.D. Student with the Department of Electrical Engineering, Columbia University, New York, from 2020 to 2021. Currently, she is an Associate Professor with the School of Electronics Engineering, Xidian University. Her research interests include signal processing, new concept radar and intelligent sensing. 
\end{IEEEbiography}

\begin{IEEEbiography}[{\includegraphics[width=1in,height=1.25in,clip,keepaspectratio]{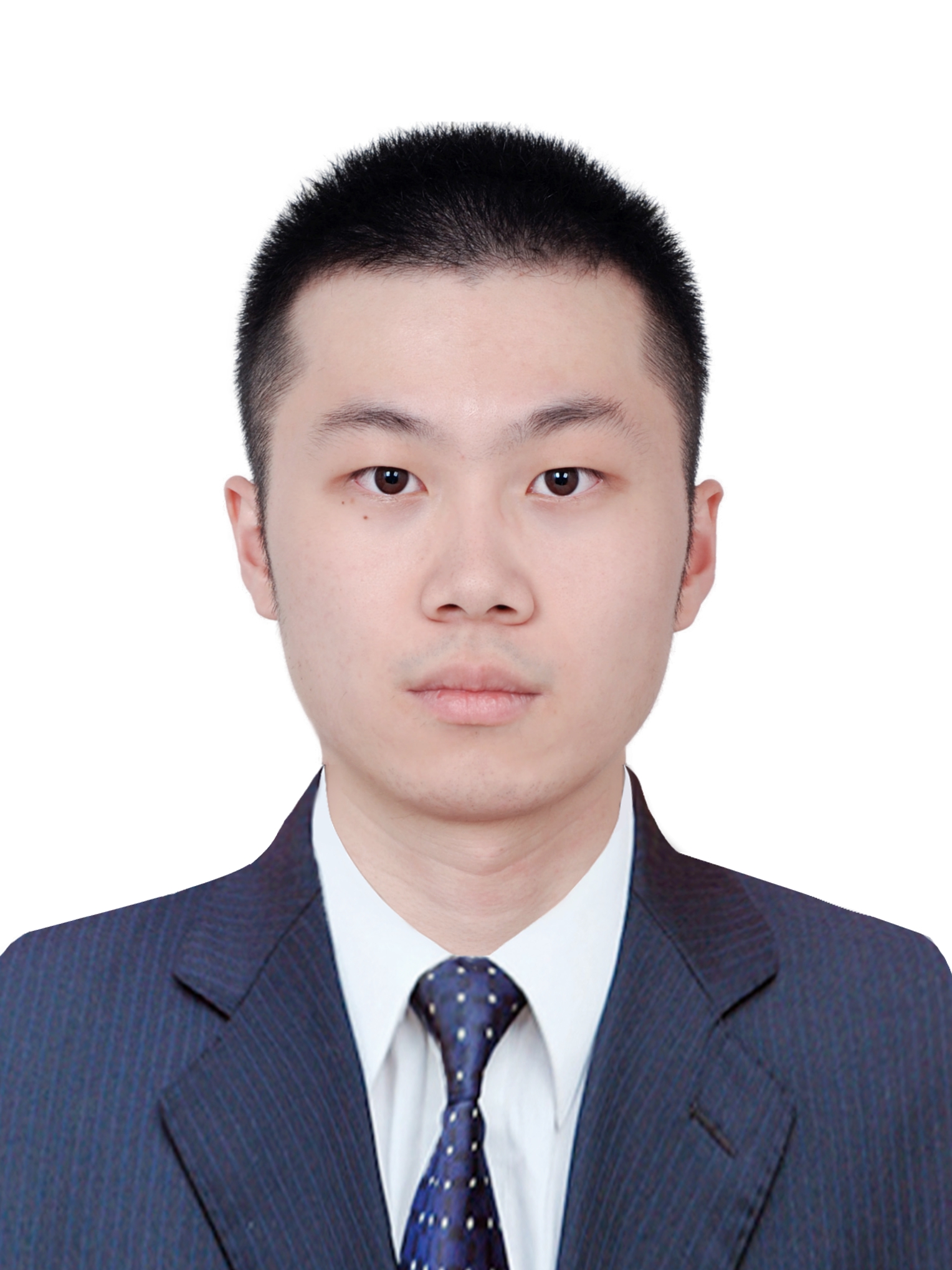}}]{Bairui Cai}
received the bachelor’s degree from Hangzhou Dianzi University, Hangzhou, China, in 2023. He is currently pursuing the master’s degree at Xidian University, Xi'an, China. 
He has issued a patent and chaired a graduate Student Innovation fund.
His research interests include radar signal processing and deep learning. 
\end{IEEEbiography}

\begin{IEEEbiography}[{\includegraphics[width=1in,height=1.25in,clip,keepaspectratio]{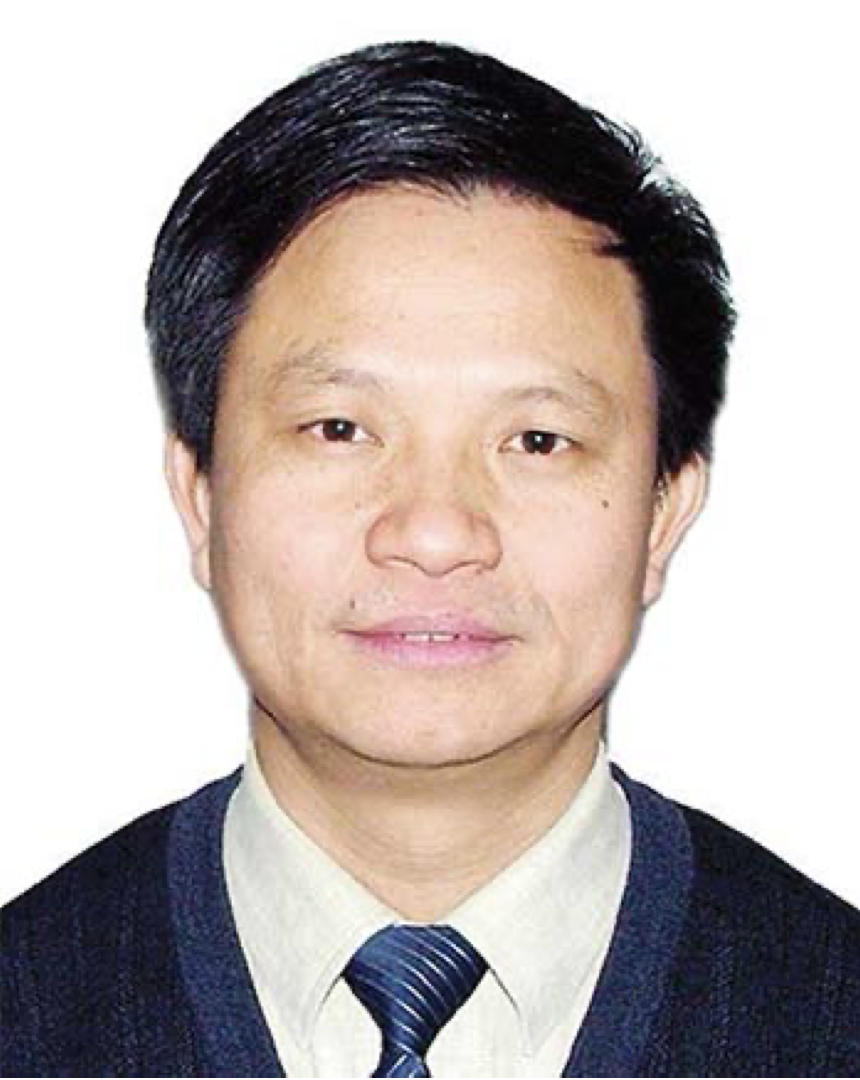}}]{Guisheng Liao}
(Senior Member, IEEE) was born in Guilin, Guangxi, China in 1963. He received the B.S. degree from Guangxi University in mathematics, Guangxi, China, in 1985, and the M.S. degree in computer software and the Ph.D. degree in signal and information processing from Xidian University, Xi’an, China, in 1990, and 1992, respectively. 

He is currently a Full Professor with the National Key Laboratory of Radar Signal Processing and served as the 1st Dean with the Hangzhou Institute of Technology, Xidian University since 2021. He has
been the Dean with the School of Electronic Engineering, Xidian University from 2013 to 2021. He has been a Senior Visiting Scholar with the Chinese University of Hong Kong from 1999 to 2000. He won the National Science Fund for Distinguished Young Scholars in 2008. His research interests include array signal processing, space-time adaptive processing, radar waveform design, and airborne/space surveillance and warning radar systems.
\end{IEEEbiography}

\vspace{11pt}

\vfill

\end{document}